\documentclass[iop,apj]{emulateapj}
\usepackage{graphicx}
\usepackage{amssymb}
\usepackage{amsmath}
\usepackage{url}

\shortauthors{Frederiks et al.}

\begin{document}

\def \GRB {GRB~110918A}

\def \KW  {Konus-\textit{WIND}}
\def \SW  {\textit{Swift}}
\def \BAT {\textit{Swift}-BAT}
\newcommand{\XRT}{\textit{Swift}-XRT }
\newcommand{\UVOT}{\textit{Swift}-UVOT }
\newcommand{\INT}{\textit{INTEGRAL} }
\newcommand{\SPIACS}{SPI-ACS }
\newcommand{\MO}{\textit{Mars~Odyssey} }
\newcommand{\MESS}{\textit{MESSENGER} }
\newcommand{\NHunits}{\mbox{$10^{21}~{\rm cm}^{-2}$}}
\newcommand{\flux}{erg cm$^{-2}$ s$^{-1}$}

\slugcomment{Submitted to ApJ 2013-09-09, revised 2013-10-28, accepted 2013-10-29}

\title{The ultraluminous GRB~110918A}

\author{
D.~D.~Frederiks\altaffilmark{1}, K.~Hurley\altaffilmark{2}, D.~S.~Svinkin\altaffilmark{1},
V.~D.~Pal'shin\altaffilmark{1}, V.~Mangano\altaffilmark{3,4}, S.~Oates\altaffilmark{5},
R.~L.~Aptekar\altaffilmark{1}, S.~V.~Golenetskii\altaffilmark{1},
E.~P.~Mazets\altaffilmark{1,13}, Ph.~P.~Oleynik\altaffilmark{1}, A.~E.~Tsvetkova\altaffilmark{1},
M.~V.~Ulanov\altaffilmark{1}, A.~V.~Kokomov\altaffilmark{1}, T.~L.~Cline\altaffilmark{6,14},
D.~N.~Burrows\altaffilmark{3}, H.~A.~Krimm\altaffilmark{6}, C.~Pagani\altaffilmark{7}, B.~Sbarufatti\altaffilmark{3,8}, M.~H.~Siegel\altaffilmark{3},
I.~G.~Mitrofanov\altaffilmark{9}, D.~Golovin\altaffilmark{9}, M.~L.~Litvak\altaffilmark{9}, A.~B.~Sanin\altaffilmark{9},
W.~Boynton\altaffilmark{10}, C.~Fellows\altaffilmark{10}, K.~Harshman\altaffilmark{10}, H.~Enos\altaffilmark{10}, R.~Starr\altaffilmark{6},
A.~von~Kienlin\altaffilmark{11}, A.~Rau\altaffilmark{11}, X.~Zhang\altaffilmark{11},
J.~Goldstein\altaffilmark{12}}
%
\altaffiltext{1}{Ioffe Physical-Technical Institute, Politekhnicheskaya 26, St.~Petersburg 194021, Russia; fred@mail.ioffe.ru }
\altaffiltext{2}{Space Sciences Laboratory, University of California, 7 Gauss Way, Berkeley, CA 94720-7450, USA}
\altaffiltext{3}{Pennsylvania State University, Department of Astronomy and Astrophysics, College Park, PA~16801, USA}
\altaffiltext{4}{INAF-IASFPA, Via Ugo La Malfa 153, 90146 Palermo, Italy}
\altaffiltext{5}{Mullard Space Science Laboratory, University College London, Holmbury St. Mary, Dorking, Surrey~RH5~6NT, UK}
\altaffiltext{6}{NASA Goddard Space Flight Center, Greenbelt, MD~20771, USA}
\altaffiltext{7}{U. Leicester, University Road, Leicester, LE1 7RH, UK}
\altaffiltext{8}{INAF-OAB, via Bianchi 46, 23807, Merate (LC), Italy}
\altaffiltext{9}{Space Research Institute, Profsoyuznaya 84/32, Moscow 117997, Russia}
\altaffiltext{10}{Department of Planetary Sciences, University of Arizona, Tucson, AZ~85721, USA}
\altaffiltext{11}{Max-Planck-Institut f\"{u}r extraterrestrische Physik, Giessenbachstrasse, Postfach 1312, D-85748 Garching, Germany}
\altaffiltext{12}{Applied Physics Laboratory, Johns Hopkins University, Laurel, MD~20723, USA}
\altaffiltext{13}{Deceased}
\altaffiltext{14}{Emeritus}

\keywords{gamma-ray burst: individual (GRB~110918A)}

\newpage
\begin{abstract}

\GRB\, is the brightest long $\gamma$-ray burst (GRB) detected by \KW\, during
its almost 19 years of continuous observations and the most luminous GRB ever observed
since the beginning of the cosmological era in 1997.
We report on the final Interplanetary Network localization of this event and
its detailed multi-wavelength study with a number of space-based instruments.
The prompt emission is characterized by a typical duration, a moderate peak energy
of the time-integrated spectrum, and strong hard-to-soft evolution.
The high observed energy fluence yields, at z=0.984, a huge isotropic-equivalent energy release
$E_{\mathrm{iso}}=(2.1\pm0.1)\times10^{54}$~erg. The record-breaking energy flux
observed at the peak of the short, bright, hard initial pulse results in an unprecedented
isotropic-equivalent luminosity $L_{\mathrm{iso}}=(4.7\pm0.2)\times10^{54}$erg~s$^{-1}$.
A tail of the soft $\gamma$-ray emission was detected with temporal and spectral behavior
typical of that predicted by the synchrotron forward-shock model.
\SW/XRT and \SW/UVOT observed the bright afterglow from 1.2 to 48 days after the burst
and revealed no evidence of a jet break. The post-break scenario for the afterglow
is preferred from our analysis, with a hard underlying electron spectrum and
ISM-like circum-burst environment implied.
We conclude that, among multiple reasons investigated, the tight collimation of the jet must
have been a key ingredient to produce this unusually bright burst.
The inferred jet opening angle of 1.7$\arcdeg$-3.4$\arcdeg$ results in reasonable
values of the collimation-corrected radiated energy and the peak luminosity, which,
however, are still at the top of their distributions for such tightly collimated events.
We estimate a detection horizon for a similar ultraluminous GRB of z$\sim$7.5 for \KW\,
and z$\sim$12 for \SW/BAT, which stresses the importance of $\gamma$-ray bursts as probes
of the early Universe.
\end{abstract}

\newpage

\section{INTRODUCTION}

Gamma-ray bursts (GRBs) are the brightest electromagnetic events known
to occur in the Universe. The bursts last from a fraction of a second
to several thousand seconds, showing a wide range of structures in their light curves
and having a typical peak energy in the 100~keV--1~MeV range.
The overall observed GRB fluences range from $10^{-7}$ to as high as $10^{-3}$~erg~cm$^{-2}$.
With the discovery of the cosmological nature of the phenomenon in 1997 \citep{Metzger1997},
it became clear that the observed flux corresponds to an enormous
isotropic luminosity, making GRBs the most luminous objects in the sky.
Out of the hundreds of GRBs so far observed with known redshifts,
there are about a dozen with an isotropic energy release $E_{\mathrm{iso}}\gtrsim10^{54}$~erg;
some of the most energetic events, such as GRB~080916C, seemingly released enough
energy in the prompt $\gamma$-rays ($E_{\mathrm{iso}} = 8.8 \times 10^{54}$ ergs, at z=4.35;
\cite{Abdo2009}; \cite{Greiner2009}) to constitute several times
the Solar rest-mass equivalent.

The hypothesis that GRBs are non-spherical explosions implies that,
when the tightly collimated relativistic fireball is decelerated
by the circum-burst medium down to the Lorentz factor $\Gamma\sim1/\theta_{\mathrm{jet}}$
(where $\theta_{\mathrm{jet}}$ is the jet opening angle), an achromatic break
(jet break) should appear, in the form of a sudden steepening in the GRB afterglow
light curve, at a characteristic time $t_{\mathrm{jet}}$.
The steepening is caused by the combination of the jet edge effect,
and the possible relativistic sideways expansion effect
\citep{Rhoads1999,SPH1999,PM1999,ZM2004}.
In the canonical light curve of X-ray afterglows
\citep{Zhang2006,Nousek2006}
the jet break corresponds to a transition from the ``normal''
segment~III to the post-break, ``jet'' segment~IV.
With typical collimation angles of a few degrees, the true energy release
from most GRBs is $\sim10^{51}$ ergs, on par with that of a supernova \citep{Frail2001}.

\begin{figure}[t!]
\centering
\includegraphics[width=0.5\textwidth]{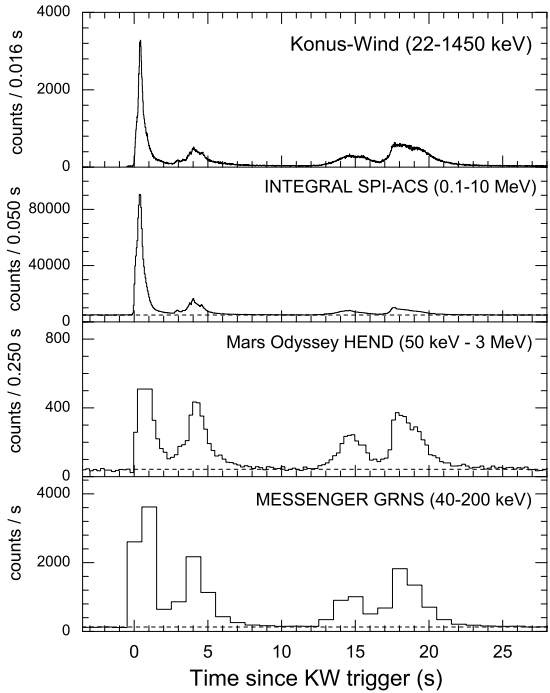}
\caption{\GRB\, light curves recorded by the four IPN instruments.
The time scale is corrected for the burst propagation between the spacecraft.
The \KW\, trigger time corresponds to the Earth-crossing time 77218.928~s UT (21:26:58.928)
}
\label{FigIPNlcs}
\end{figure}
\begin{figure}[t!]
\centering
\includegraphics[width=0.5\textwidth]{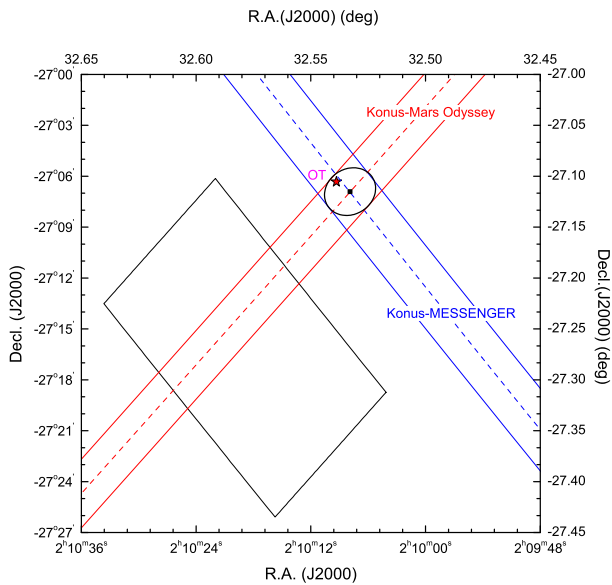}
\caption{Initial IPN error box of \GRB, the refined annuli,
the error ellipse, and the X-ray/optical counterpart (indicated by the star).}
\label{FigIPNmap}
\end{figure}
The long-duration, extremely intense \GRB\, was detected
by \emph{INTEGRAL} (SPI-ACS), \KW, \emph{Mars~Odyssey} (HEND),
and \emph{MESSENGER} (GRNS). At the time of the burst, \SW\,
was in the South Atlantic Anomaly and Earth-occulted;
\emph{Fermi} was also Earth-occulted.
Using the Interplanetary Network (IPN), a location was determined \citep{GCN12357}.
A preliminary analysis of the \KW\, detection revealed that GRB~110918A
is the most intense $\gamma$-ray burst observed by the instrument
since it began operation in 1994 November \citep{GCN12362}.

\SW/XRT began follow-up observations $\sim$1.2~days after the trigger and
was able to observe and localize an X-ray counterpart to this burst \citep{GCN12364}.
The XRT source was found outside the $3\sigma$ initial IPN error box at 640$\arcsec$ from its center.
The optical afterglow \citep{GCN12365} was monitored by \SW/UVOT \citep{GCN12371}
and by a multitude of ground-based telescopes \citep{GCN12366,GCN12367,GCN12368,GCN12369,GCN12382,GCN12388,Elliott2013}.
No sub-mm flux was detected from the source, with a 3-sigma upper limit of 15~mJy,
in the APEX/LABOCA observations performed at 345~GHz \citep{GCN12381}.
The best optical counterpart position is from GROND \citep{Elliott2013}
at a location R.A.(J2000) = $02^h10^m09.34^s$, Decl.(J2000) = $-27\arcdeg06\arcmin19.7\arcsec$ with an error of 0.2$\arcsec$.
Using the GMOS-N spectrograph on Gemini-N, Mauna Kea, \cite{GCN12368}
determined a spectroscopic redshift of $z=0.982$. This was later confirmed by
\cite{GCN12375} with the GTC telescope at Roque de los Muchachos
Observatory, who reported $z=0.984\pm0.001$.

At this redshift, the huge energy flux measured by \KW\, implied equally
enormous values of the isotropic-equivalent energy released in the source frame,
$E_{\mathrm{iso}}\sim1.9\times10^{54}$~erg, and an isotropic-equivalent peak luminosity
$L_{\mathrm{iso}}\sim4.4\times10^{54}$erg~s$^{-1}$ \citep{GCN12370}.
These preliminary estimates place \GRB\, among the several brightest events
ever observed in the era of cosmological GRBs and, accordingly, this burst merits
a more detailed consideration.

The main goal of this article is to give a comprehensive coverage of \GRB\, observations
made with space-based instruments. In Section~\ref{sec:ipn},
we start with an analysis of the IPN localization, discuss the spacecraft timing issues
and provide corrected annuli, which form a refined error ellipse.
In Sections~\ref{sec:kw} and \ref{sec:extended}, we give a detailed description
and analysis of the prompt and extended $\gamma$-ray emission detected by \KW.
In Section~\ref{sec:afterglow}, we report on results of X-ray afterglow
observations made with \SW-XRT and UV/optical observations with \SW-UVOT.
Finally, we discuss these results in the context of the burst cosmological rest frame
and put constraints on the collimation angle and the collimation-corrected energy of \GRB.
In a companion paper, \citep{Elliott2013}, hereafter E13, discuss the ground-based
optical observations of this event and its host galaxy.

Throughout the paper all errors reported are 90\% conf. levels unless otherwise specified.
The power-law temporal decay slopes, $\alpha$, and power-law spectral indices, $\beta$,
are defined such that the flux density $F_{\nu}(t)\propto t^{-\alpha}$
and $F_{\nu}\propto \nu^{-\beta}$, respectively; also, to avoid a confusion
with the Band spectral model parameters, the use of $\alpha$ and $\beta$
is explicitly stated where appropriate. We adopt the conventional notation $Q_k=Q/10^k$,
and use cgs units unless otherwise noted.


\section{OBSERVATIONS AND ANALYSES}

\subsection{IPN Observations and Localization}
\label{sec:ipn}

\begin{figure}[t!]
\centering
\includegraphics[width=0.5\textwidth]{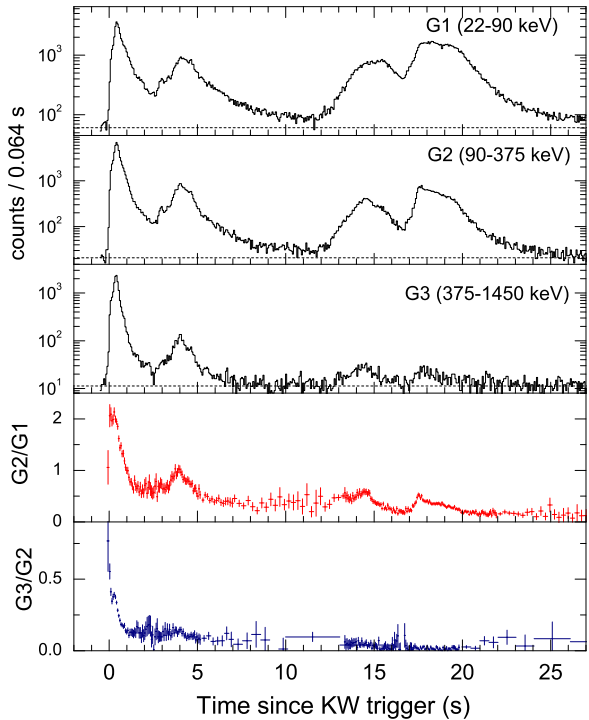}
\caption{Light curves of \GRB\, recorded by \KW\, in the G1, G2 and G3 energy
bands with 64~ms resolution (three upper panels).
The count rates are dead-time corrected; background levels are indicated by dashed lines.
An apparent hardness-intensity correlation and a general emission softening
in the course of the burst are illustrated by the evolution of the hardness ratios
(G2/G1 and G3/G2) shown in the two lower panels.
}
\label{FigKW_64ms_G1G2G3log}
\end{figure}

\GRB\, was detected by four IPN: \emph{INTEGRAL} SPI-ACS \citep{Rau2005}, in a highly elliptical orbit,
\KW\, \citep{Aptekar1995}, in orbit around the Lagrangian point L1, \emph{MESSENGER} GRNS \citep{Gold2001}, in orbit around Mercury, and \emph{Mars~Odyssey} HEND \citep{Hurley2006},
in orbit around Mars, at 0.46, 5.0, 645.9, and 943.0 light-seconds from Earth, respectively.
The light curve of the event (Figure~\ref{FigIPNlcs}) starts with an extremely bright, hard, and short pulse followed
by three weaker, softer, and partly overlapping pulses in the next 25 seconds.

An initial 62~sq.~arcmin IPN error box was derived using \KW, \emph{MESSENGER}, and \emph{Odyssey},
and was announced in a GCN Circular \citep{GCN12357}. A \SW\, target-of-opportunity observation was requested; when the XRT
position was announced \citep{GCN12364} it was evident that the error box was significantly displaced
from the counterpart.  After lengthy investigation, it was found that both the \emph{MESSENGER} and \emph{Odyssey}
times were inaccurate due to the use of outdated spacecraft clock files.  With the updated files, and with
\emph{INTEGRAL} added, the burst was triangulated again, and an error ellipse was derived using the method described in \cite{Hurley2000}.
The 3$\sigma$ ellipse has major axis 3.16~arcmin, minor axis 1.2~arcmin, and area 2.6~sq.~arcmin;
the ellipse is centered at R.A.(J2000)=$02^h10^m07.9s$, Decl.(J2000)=$-27\arcdeg06\arcmin54.4\arcsec$, with $\chi^2 =0.06$ for
1~degree of freedom (d.o.f.; three annuli minus two fitted coordinates).
The optical counterpart found by \citet{GCN12365} lies inside the ellipse, 0.66~arcmin from its center (Figure~\ref{FigIPNmap}).

\subsection{\KW\, Observation and Analysis}
\label{sec:kw}

\GRB\, triggered the \KW\, $\gamma$-ray spectrometer (KW)
at $T_0(KW)$=77222.856~s UT (21:27:02.856) on 2011 September 18, hereafter $T_0$.
It was detected by the S1 detector, which observes the Southern ecliptic hemisphere;
the incident angle was $53\fdg1$.
The propagation delay from Earth to \emph{WIND} is 3.928 s for this GRB, 
correcting for this factor, the KW trigger time corresponds
to the Earth-crossing time 77218.928~s UT (21:26:58.928).
\begin{figure}
\centering
\includegraphics[width=0.5\textwidth]{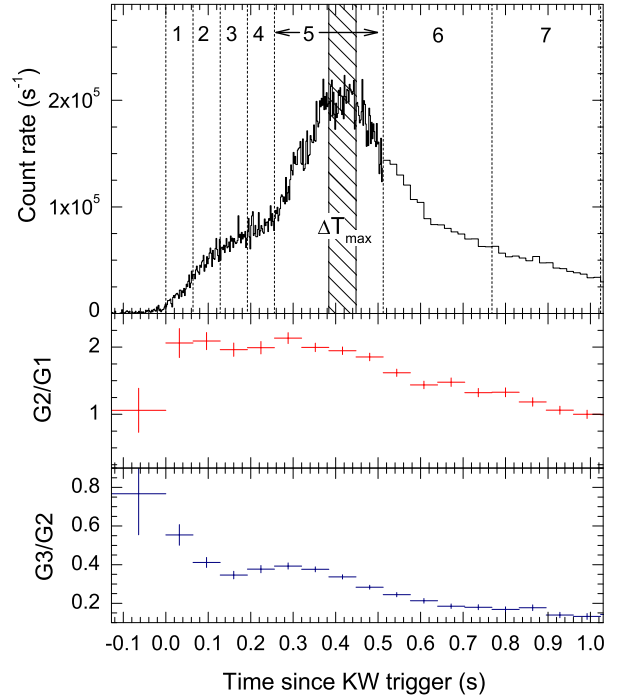}
\caption{Initial pulse, $P_1$. \KW\, light curve in the G1+G2+G3 (22--1450~keV) band is shown with
a 2~ms (16~ms after $T_0+0.512$~s) resolution.
Accumulation intervals for the KW time-resolved spectra 1--7 are indicated with the vertical dashed lines.
The 64~ms interval $\Delta T_{\mathrm{max}}$ has been used for the peak energy flux calculation.
The hardness ratios are shown in the two lower panels.}
\label{FigKWlc2}
\end{figure}

\subsubsection{Time History}\label{sec:timehist}

In the instrument's ``triggered mode'', count rates are recorded in three energy bands:
G1(22--90 keV), G2(90--375 keV), and G3(375--1450 keV). The record starts
at $T_0-0.512$~s and continues to $T_0+229.376$~s with an accumulation time varying from 2 to 256~ms.
The ``background mode'' count rate data are available up to $T_0+250$~s in the same energy bands with
a coarse resolution of 2.944~s.

The prompt-emission light curve (Figure~\ref{FigKW_64ms_G1G2G3log}) can be divided into
two groups of overlapping pulses, which are separated by a minimum around $\sim~T_0+11.5$~s
when the observed rate in the harder G2 and G3 bands is comparable to the background level.
The first group, hereafter referred as Phase~I, is characterized by two pronounced
pulses: the huge $P_1$, which peaks at $\sim~T_0+0.368$~s,
and $P_2$ ($\sim~T_0+4.032$~s); the second group (Phase~II)
is comprised of considerably overlapping pulses $P_3$ ($\sim~T_0+14.6$~s) and $P_4$ ($\sim~T_0+17.7$~s).
Shown in the same figure, the temporal evolution of the G2/G1 and G3/G2 hardness
ratios indicates an apparent hardness-intensity correlation of the emission
against a general tendency of spectral softening in the course of the burst.

The bright, pulsed emission decays up to $\sim~T_0+30$~s.
However, a stable excess in the count rate over the background level was detected,
mostly in the softer G1 and G2 energy bands, until $T_0+250$~s when the KW
measurements stopped due to the limited capacity of the {\it WIND} spacecraft telemetry.
The detailed analysis of the extended $\gamma$-ray emission is given in Section~\ref{sec:extended}.

The burst starts with a sharp rise of the bright, hard, short pulse
$P_1$ which culminates after a $\sim$350~ms two-step onset (Figure~\ref{FigKWlc2}).
In this phase, a spectrum with high and evolving peak energy is suggested by the hardness
ratio behavior: the G3/G2 ratio rapidly decays from a maximum at the onset of
the initial pulse, while G2/G1 remains at a high, but stable level until $\sim~T_0+0.370$~s,
when a peak of the emission is reached.
The observed count rate of $\sim~2\times 10^5$~s$^{-1}$ in the cumulative G1+G2+G3 (22--1450~keV)
energy band is unprecedented in almost 19 years of KW observations
of long $\gamma$-ray bursts.
Although the photon flux is very high, a standard KW dead-time
correction procedure (i.e., a simple non-paralyzable dead-time
correction in each of the measurement bands, taking into account
a softer gate blocking by harder ones) is still applicable to the burst;
no additional modeling, which was used, e.g., in an analysis
of the KW detection of the 1998 August 27 giant flare from SGR~1900+14,
is required (details of these simulations and
the KW dead-time correction procedures can be found in \citealt{Mazets1999}).

The trailing edge of the initial pulse is more gently sloping;
the emission intensity decreases until $T_0+2.5$~s and gives way,
with a small ``bump'' at $\sim~T_0+2.95$~s,
to the rise of the second, relatively weaker, and softer pulse, $P_2$.
The G1+G2+G3 count rate in this pulse reaches a 64~ms peak value
of $\sim3\times 10^4$~s$^{-1}$ (or $\sim0.15$ of that in $P_1$)
at $T_0+4.032$~s and then gradually decays to a minimum around $\sim~T_0+11.5$~s,
which separates Phase~I and Phase~II.

A new rise begins at $\sim~T_0+11.5$~s. The third ($P_3$) and fourth ($P_4$)
pulses form an overlapping structure (Phase~II) in the time interval from $T_0+11.5$~s to $\sim~T_0+25$~s.
The peak 64-ms count rate reached in these two pulses is $\sim1.9\times10^4$~s$^{-1}$
and $\sim3.8\times 10^4$~s$^{-1}$, respectively.
While these rates are on par with that in the second pulse, $P_2$,
the hardness ratios indicate a considerably softer emission spectrum.
Despite the huge count rate in the initial pulse, the count fluence recorded
by KW in the 22--1450~keV band during the first $\sim25$ seconds
of the burst is nearly equally divided between Phase~I and Phase~II.

In the G2+G3 energy band (90--1450 keV), standard for calculations of the KW
GRB light curve characteristics, the total duration of the burst,
determined at the 5$\sigma$ level, is
$T_{100}=95.154$~s (from $T_0-0.178$~s to $T_0+94.976$~s).
The corresponding $T_{90}$ value is $19.6\pm0.1$~s and $T_{50}=14.3\pm0.1$~s.
The G2+G3 count rate reached a maximum of $(1.46\pm 0.02)\times 10^5$~s$^{-1}$
in the 64~ms bin starting 0.368~s after the trigger, and the total
number of counts during the $T_{100}$ interval is $1.34\times10^5$.

\begin{deluxetable}{lcccc}[t!]
\tablewidth{0pt}
\tablecaption{Spectral Lags between \KW\, Light Curves
\label{TableLag}}
\tablehead{
\colhead{Time Interval} & \colhead{Light Curves} & \colhead{$\tau_{\mathrm{lag}}\tablenotemark{a}$}\\
\colhead{from $T_0$(s)} & \colhead{} & \colhead{(s)}
}
\startdata
-0.032--26.368  & G3--G1    & 0.092$\pm$0.003\tablenotemark{b} \\
                & G2--G1    & 0.047$\pm$0.002 \\
                & G3--G2    & 0.047$\pm$0.002 \\
\\
-0.032--2.000   & G3--G1    & 0.091$\pm$0.003 \\
(Initial pulse) & G2--G1    & 0.043$\pm$0.002 \\
                & G3--G2    & 0.047$\pm$0.002 \\
\\
-0.032--8.000   & G3--G1    & 0.092$\pm$0.003 \\
(Phase~I)       & G2--G1    & 0.044$\pm$0.002 \\
                & G3--G2    & 0.047$\pm$0.002 \\
\\
12.096--26.368  & G3--G1    & 0.36$\pm$0.13 \\
(Phase~II)      & G2--G1    & 0.19$\pm$0.01 \\
                & G3--G2    & 0.09$\pm$0.05 \\
\enddata
\tablenotetext{a}{\footnotesize{Positive spectral lag $\tau_{\mathrm{lag}}$ means that the spectrum has hard to soft evolution.}}
\tablenotetext{b}{\footnotesize{1$\sigma$ uncertainties throughout this table.}}
\end{deluxetable}

\subsubsection{Spectral Lags}
The observed evolution of the hardness ratios suggests that
soft photons are delayed with respect to the higher energy ones,
a common property of long GRBs.
We examined the spectral lag $\tau_{\mathrm{lag}}$ using the cross-correlation function
(CCF) between the light curves in two energy bands \citep{Norris2000,Band1997}.
After calculating the CCF as a function of $\tau_{\mathrm{lag}}$ we obtained
the peak value of $\tau_{\mathrm{lag}}$ by fitting it with a fourth-degree polynomial.

The resulting values of $\tau_{\mathrm{lag}}$ between the 16~ms G1, G2, and G3 light curves
at different phases of the burst are listed in Table~\ref{TableLag}.
Statistically significant lags in the $40\,-\,360$~ms range are derived for
the initial pulse ($P_1$), Phase~I, Phase~II, and in the entire prompt phase of the emission.
The positive $\tau_{\mathrm{lag}}$'s are indicative of hard-to-soft spectral evolution.

Predictably, the lags for the entire burst and Phase~I
are almost identical to the lags found for the huge initial pulse alone.
However, for the soft Phase~II, the values of $\tau_{\mathrm{lag}}$
are two-to-four times longer than for the hard initial phase of the event.

\subsubsection{Light-curve Decomposition} \label{sec:pulses}

\begin{figure}
\centering
\includegraphics[width=0.5\textwidth]{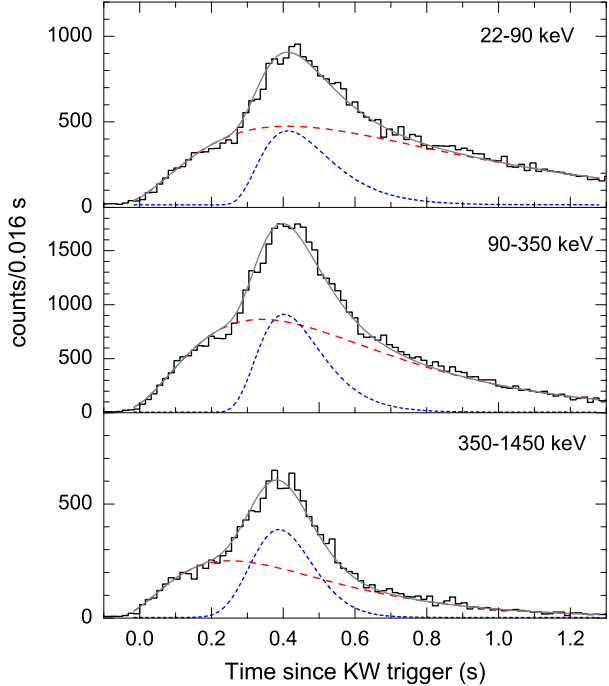}
\caption{Decomposition of the initial pulse light curve.
The background-subtracted 16~ms light curves in the G1, G2, and G3 bands
are best fit by two components, each described by the four-parameter
exponential model. The components $C_1$ and $C_2$ are shown with long
and short dashes, respectively; the sum of the components is shown with a solid line.
}
\label{FigPulses}
\end{figure}

The prompt emission of long GRBs shows a wide range of structures in its light curve.
Attempts have been made to fit the light curves with Fast Rise Exponential Decay
models described by combinations of power-law functions \citep{KL2003,KRL2003},
exponential functions \citep{Norris1996,LBP2000a,LBP2000b,Norris2005}, and log-normal functions \citep{Bhat2012}.

The light curve of the initial pulse (Figure~\ref{FigKWlc2}) shows a clear two-step onset,
which suggests at least two overlapping episodes of emission.
We performed the analysis of this pulse with the background-subtracted
G1, G2, and G3 light curves using the four-parameter exponential model from \cite{Norris2005}:
\begin{equation}
A(t)=A_m\lambda \exp\{-\tau_{1}/(t-t_0)-(t-t_0)/\tau_{2}\}\label{PulseExp}
\end{equation}
for $t>t_0$, where $\lambda=\exp\left(2\mu\right)$, $\mu=\left(\tau_{1}/\tau_{2}\right)^{1/2}$.
$A_m$ is the pulse amplitude, $t_0$ is the pulse start time, and $\tau_{1}$, $\tau_{2}$
are time constants characterizing the rise and decay parts of the pulse.
This pulse peaks at $t_m=t_0+\left(\tau_{1}\tau_{2}\right)^{1/2}$ and
has a width measured between two $1/e$ points, $w=\tau_2(1+4\mu)^{1/2}$.

The 16~ms light curves have been fitted in the $T_0\,-\,T_0+1.3$~s interval
with a single pulse, two, and more overlapping components using a $\chi^2$ statistic.
The single-component model yields very high $\chi^2_r$.
Introducing the second component improved the fit dramatically:
$\chi^2$ changed from 1123/78~d.o.f. to 132/74~d.o.f. for the G2 light curve.
A similar improvement is found for the G1 and G3 light curves.
Further addition of components to the model results
in ambiguous, poorly constrained fits and does not improve the statistic.
Thus, we conclude that the KW light curves of the initial
pulse are best described by the double-component model.

\begin{deluxetable*}{ccccccc}
\tabletypesize{\scriptsize}
\tablewidth{0pt}
\tablecaption{Initial Pulse Decomposition
\label{TableNorrisPulses}}
\tablehead{
\colhead{Light} & \colhead{$A_{m}$} & \colhead{$t_0$} & \colhead{$\tau_1$} & \colhead{$\tau_2$} & \colhead{$t_m$}  & \colhead{$w$} \\
\colhead{curve} & \colhead{($10^3$~counts~s$^{-1}$)} & \colhead{(s)} & \colhead{(s)} & \colhead{(s)} & \colhead{(s)} & \colhead{(s)}}
\startdata
\multicolumn{6}{c}{Component $C_1$} \\[0cm]
G1&   28.7 $\pm$ 0.3 & -0.144 $\pm$ 0.002 &  0.631 $\pm$ 0.009 & 0.491 $\pm$ 0.005 & 0.412 $\pm$ 0.007 & 1.154 $\pm$ 0.012\\
G2&   53.7 $\pm$ 0.4 & -0.157 $\pm$ 0.001 &  0.801 $\pm$ 0.007 & 0.307 $\pm$ 0.002 & 0.339 $\pm$ 0.005 & 0.839 $\pm$ 0.005\\
G3&   15.5 $\pm$ 0.2 & -0.150 $\pm$ 0.003 &  0.648 $\pm$ 0.011 & 0.250 $\pm$ 0.003 & 0.262 $\pm$ 0.007 & 0.682 $\pm$ 0.007\\
\\
\multicolumn{6}{c}{Component $C_2$} \\[0cm]
G1&   27.1 $\pm$ 0.7 &  0.177 $\pm$ 0.003 &  0.706 $\pm$ 0.021 & 0.079 $\pm$ 0.002 & 0.413 $\pm$ 0.004 & 0.283 $\pm$ 0.006\\
G2&   56.7 $\pm$ 1.1 &  0.137 $\pm$ 0.002 &  1.221 $\pm$ 0.020 & 0.057 $\pm$ 0.001 & 0.402 $\pm$ 0.003 & 0.253 $\pm$ 0.003\\
G3&   24.1 $\pm$ 0.6 & -0.179 $\pm$ 0.002 & 12.418 $\pm$ 0.107 & 0.026 $\pm$ 0.001 & 0.388 $\pm$ 0.005 & 0.244 $\pm$ 0.002\\
\enddata
\end{deluxetable*}

Figure~\ref{FigPulses} shows the decomposition and Table~\ref{TableNorrisPulses} lists
the model parameters and the derived quantities. The narrow component $C_2$, which dominates
at the peak of the emission in the harder bands, is delayed with respect to a wider component $C_1$,
which describes the onset of the initial pulse before $\sim~T_0+0.2$~s and its decay after $\sim~T_0+0.8$~s.
The delays between the components, calculated as the difference between their peak times, $t_m$,
are $\sim0$~ms, $\sim62$~ms, and $\sim136$~ms in the G1, G2, and G3 bands, respectively.
In the same way, we calculated lags between the G1, G2, and G3 bands
for each component. The G3-G1, G2-G1, and G3-G2 lags for the leading sub-pulse $C_1$
(160$\pm$10~ms, 72$\pm$8~ms,  and 87$\pm$8~ms, respectively)
are $\sim7$ times longer than the corresponding lags for $C_2$ (24$\pm$7~ms, 10$\pm$5~ms, and 14$\pm$6~ms, respectively).
The relatively longer lags obtained for the leading component are in agreement with
the strong evolution of the G3/G2 hardness ratio present in the first $\sim$300~ms of the burst.

Since a rapid variability in GRB light curves could be linked directly to the activity
of the central engine \citep{SP97,Kobayashi1997,Ryde2004},
the shortest variability timescale for a gamma-ray burst is clearly of interest.
In particular, assuming that the shortest timescale in GRB prompt emission is
the shortest pulse width, the length scale of the GRB central engine can be estimated
(see, e.g., \cite{Bhat2012} and references therein).
For \GRB, we find a width of $\delta T\equiv w=0.25$~s,
obtained for the sub-pulse $C_2$ in the energy bands G2 and G3.

\subsubsection{Time-resolved Spectral Analysis}\label{sec:specres}

\begin{deluxetable*}{lcccccc}
\tabletypesize{\scriptsize}
\tablewidth{0pt}
\tablecaption{\KW\, Time-resolved Spectral Fits with the Band Function
\label{TableKWSpecRes}}
\tablehead{
\colhead{Spectrum} & \colhead{Accumulation} & \colhead{$\alpha$} & \colhead{$\beta$} & \colhead{$E_{\mathrm{peak}}$} & \colhead{Flux\tablenotemark{a}} & \colhead{$\chi^2/$d.o.f.}\\
\colhead{} & \colhead{Interval (s from $T_0$)} & \colhead{} & \colhead{} & \colhead{(keV)} & \colhead{($10^{-6}$erg~cm$^{-2}$~s$^{-1})$} & \colhead{}
}
\startdata
1 & 0--0.064 & $-0.75_{-0.09}^{+0.10}$ & $-2.35$\tablenotemark{b} & $4050_{-2770}^{+1610}$ & $290_{-19}^{+20}$ & $22.7/25$ \\
2 & 0.064--0.128 & $-0.67_{-0.09}^{+0.10}$ & $-2.35$\tablenotemark{b} & $1630_{-270}^{+330}$ & $348_{-20}^{+21}$ & $33.3/33$ \\
3 & 0.128--0.192 & $-0.68_{-0.11}^{+0.12}$ & $-2.49_{-0.40}^{+0.29}$ & $1230_{-260}^{+300}$ & $370_{-30}^{+32}$ & $65.3/75$ \tablenotemark{c} \\
4 & 0.192--0.256 & $-0.54_{-0.13}^{+0.18}$ & $-2.23_{-0.26}^{+0.23}$ & $1160_{-300}^{+330}$ & $487_{-37}^{+40}$ & $76.7/77$ \tablenotemark{c} \\
5 & 0.256--0.512 & $-0.47_{-0.08}^{+0.09}$ & $-2.39_{-0.14}^{+0.11}$ & $1120_{-150}^{+160}$ & $775_{-26}^{+27}$ & $92.4/86$ \tablenotemark{c} \\
6 & 0.512--0.768 & $-0.44_{-0.10}^{+0.11}$ & $-2.45_{-0.10}^{+0.09}$ & $443_{-41}^{+45}$ & $289_{-12}^{+12}$ & $77.2/57$ \\
7 & 0.768--1.024 & $-0.48_{-0.15}^{+0.17}$ & $-2.48_{-0.16}^{+0.15}$ & $290_{-32}^{+40}$ & $125_{-6.8}^{+6.8}$ & $30.3/50$ \\
8 & 1.024--1.280 & $-0.88_{-0.14}^{+0.18}$ & $-2.39_{-0.27}^{+0.23}$ & $270_{-45}^{+62}$ & $51.5_{-4.4}^{+4.5}$ & $31.2/44$ \\
9 & 1.280--1.536 & $-1.26_{-0.14}^{+0.14}$ & $-2.37_{-2.60}^{+0.22}$ & $294_{-62}^{+118}$ & $24.6_{-4.4}^{+3.7}$ & $54.2/74$ \\
10 & 1.536--2.304 & $-1.34_{-0.11}^{+0.15}$ & $-2.33_{-0.56}^{+0.16}$ & $260_{-51}^{+74}$ & $13.1_{-1.7}^{+2.0}$ & $52.7/50$ \\
11 & 2.304--3.328 & $-1.25_{-0.09}^{+0.10}$ & $-2.58_{-0.53}^{+0.24}$ & $261_{-35}^{+43}$ & $10.9_{-1.1}^{+1.2}$ & $51.1/51$ \\
12 & 3.328--3.840 & $-0.98_{-0.11}^{+0.12}$ & $-2.89_{-1.85}^{+0.32}$ & $301_{-40}^{+51}$ & $24.8_{-2.2}^{+2.1}$ & $65.5/48$ \\
13 & 3.840--4.096 & $-0.97_{-0.10}^{+0.11}$ & $-2.75_{-0.62}^{+0.28}$ & $378_{-53}^{+58}$ & $48.7_{-4.2}^{+4.3}$ & $68.5/44$ \\
14 & 4.096--4.352 & $-1.00_{-0.11}^{+0.14}$ & $-2.78_{-0.61}^{+0.31}$ & $300_{-48}^{+45}$ & $41.1_{-3.4}^{+3.7}$ & $46.6/43$ \\
15 & 4.352--4.608 & $-0.86_{-0.18}^{+0.30}$ & $-2.49_{-0.29}^{+0.19}$ & $205_{-44}^{+45}$ & $33.7_{-3.0}^{+3.3}$ & $36.6/41$ \\
16 & 4.608--4.864 & $-0.69_{-0.24}^{+0.28}$ & $-2.29_{-0.15}^{+0.11}$ & $155_{-23}^{+31}$ & $29.2_{-2.8}^{+3.0}$ & $40.7/39$ \\
17 & 4.864--5.632 & $-1.28_{-0.13}^{+0.16}$ & $-2.44_{-0.28}^{+0.16}$ & $150_{-22}^{+26}$ & $12.1_{-1.2}^{+1.2}$ & $38.9/47$ \\
18 & 5.632--13.312 & $-1.33_{-0.14}^{+0.18}$ & $-2.60_{-0.27}^{+0.17}$ & $90_{-9}^{+9}$ & $1.94_{-0.16}^{+0.18}$ & $63.8/82$ \\
19 & 13.312--14.080 & $-0.79_{-0.20}^{+0.25}$ & $-2.68_{-0.20}^{+0.15}$ & $99_{-9}^{+9}$ & $9.69_{-0.66}^{+0.73}$ & $29.5/44$ \\
20 & 14.080--14.336 & $-0.60_{-0.28}^{+0.36}$ & $-2.68_{-0.31}^{+0.20}$ & $116_{-15}^{+16}$ & $16.2_{-1.5}^{+1.7}$ & $28.0/34$ \\
21 & 14.336--14.592 & $-0.64_{-0.21}^{+0.26}$ & $-2.99_{-0.43}^{+0.26}$ & $119_{-11}^{+11}$ & $17.5_{-1.3}^{+1.5}$ & $28.8/33$ \\
22 & 14.592--14.848 & $-0.87_{-0.22}^{+0.24}$ & $-3.28_{-2.30}^{+0.38}$ & $111_{-10}^{+12}$ & $15.8_{-1.2}^{+1.3}$ & $29.1/33$ \\
23 & 14.848--15.360 & $-0.96_{-0.15}^{+0.18}$ & $-3.58_{-0.83}^{+0.39}$ & $85_{-5}^{+5}$ & $12.9_{-0.58}^{+0.63}$ & $26.0/38$ \\
24 & 15.360--16.128 & $-1.12_{-0.21}^{+0.26}$ & $-3.11_{-0.29}^{+0.19}$ & $59_{-3}^{+3}$ & $9.99_{-0.42}^{+0.45}$ & $37.0/39$ \\
25 & 16.128--17.664 & $-1.39_{-0.10}^{+0.10}$ & $<$~-3.39               & $68_{-4}^{+3}$ & $9.63_{-0.33}^{+0.32}$ & $37.8/51$ \\
26 & 17.664--17.920 & $-0.73_{-0.22}^{+0.28}$ & $-3.68_{-1.35}^{+0.46}$ & $95_{-6}^{+6}$ & $26.3_{-1.4}^{+1.5}$ & $37.7/33$ \\
27 & 17.920--18.176 & $-0.64_{-0.18}^{+0.25}$ & $<$~-3.70               & $87_{-5}^{+4}$ & $23.2_{-0.95}^{+1.0}$ & $33.9/31$ \\
28 & 18.176--18.432 & $-0.73_{-0.20}^{+0.28}$ & $-3.78_{-0.83}^{+0.43}$ & $81_{-5}^{+5}$ & $23.9_{-1.00}^{+1.0}$ & $42.6/30$ \\
29 & 18.432--18.688 & $-0.66_{-0.22}^{+0.24}$ & $-3.65_{-0.73}^{+0.34}$ & $82_{-4}^{+5}$ & $22.9_{-1.0}^{+1.0}$ & $32.9/29$ \\
30 & 18.688--18.944 & $-0.81_{-0.22}^{+0.23}$ & $<$~-3.75               & $81_{-5}^{+5}$ & $20.9_{-0.91}^{+0.92}$ & $16.6/27$ \\
31 & 18.944--19.200 & $-0.67_{-0.21}^{+0.21}$ & $<$~-4.01               & $75_{-3}^{+4}$ & $19.4_{-0.82}^{+0.78}$ & $33.9/29$ \\
32 & 19.200--19.456 & $-0.62_{-0.29}^{+0.32}$ & $-3.82_{-2.04}^{+0.40}$ & $71_{-4}^{+4}$ & $19.1_{-0.82}^{+0.84}$ & $21.3/29$ \\
33 & 19.456--19.712 & $-0.75_{-0.28}^{+0.34}$ & $-3.67_{-0.73}^{+0.38}$ & $67_{-4}^{+4}$ & $18.1_{-0.87}^{+0.92}$ & $22.7/27$ \\
34 & 19.712--20.224 & $-0.98_{-0.22}^{+0.33}$ & $-3.63_{-0.59}^{+0.34}$ & $52_{-3}^{+3}$ & $13.0_{-0.47}^{+0.50}$ & $34.6/35$ \\
35 & 20.224--28.416 & $-1.50_{-0.31}^{+0.38}$ & $-3.11_{-0.25}^{+0.18}$ & $34_{-10}^{+8}$ & $1.46_{-0.076}^{+0.075}$ & $80.1/94$
\enddata
\tablenotetext{a}{\footnotesize{In the 20~keV--10~Mev energy band.}}\\
\tablenotetext{b}{\footnotesize{Spectra 1 and 2 were fitted with index $\beta$ fixed to that of the time-averaged spectrum 1+2,
for which: $\alpha=-0.68_{-0.08}^{+0.10}$, $\beta=-2.35_{-0.79}^{+0.32}$, and $E_{\mathrm{peak}}=2290_{-540}^{+610}$ ($\chi^2=61.6/57$~d.o.f.)}}\\
\tablenotetext{c}{\footnotesize{PG-statistic/d.o.f.}}
\end{deluxetable*}

In the triggered mode, \KW\, measures 64 energy spectra in 128
channels of two overlapping energy bands: PHA1 (22--1450~keV) and PHA2 (375~keV--18~MeV).
The first four spectra have a fixed accumulation time of 64~ms;
after that, the accumulation time varies over 0.256--8.192~s,
depending on the current intensity of the burst.
During the bright prompt phase of \GRB, 35 energy spectra were measured:
17 of them covered the first hard pair of pulses ($T_0\,-\,T_0$+5.632 s),
and spectrum 18 was recorded during the temporary decrease
in the burst intensity ($T_0+5.632\,-\,T_0+13.312$~s);
the remaining 17 spectra ($T_0+13.312\,-\,T_0+28.416$~s)
covered the second soft group of pulses and the transition to
the extended emission tail.

\begin{figure}
\centering
\includegraphics[width=0.5\textwidth]{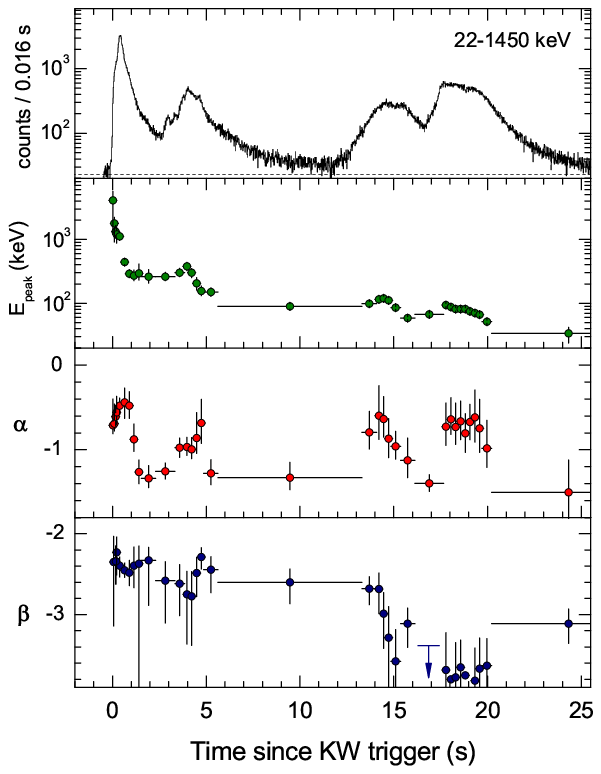}
\caption{Spectral evolution of the $\gamma$-ray emission during the
prompt phase of the burst. The \KW\, light curve in the combined G1+G2+G3 energy band (22--1450~keV)
is shown with 16~ms resolution, along with the temporal behavior
of the Band spectral model parameters $E_{\mathrm{peak}}$, $\alpha$ and $\beta$ obtained from
the time-resolved fits (see Table~\ref{TableKWSpecRes}).
}
\label{FigKWspRes}
\end{figure}

\begin{figure}
\centering
\includegraphics[width=0.5\textwidth]{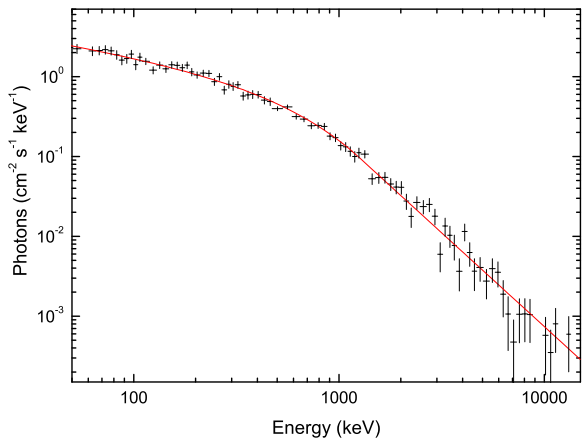}
\caption{Photon spectrum at the culmination of the initial pulse
($T_0+0.256\,-\,T_0+0.512$; the model parameters are given in Table~\ref{TableKWSpecRes}).
The emission is traced to $>10$~MeV with no obvious high-energy
cutoff observed up to the KW upper energy threshold.
}

\label{FigKWsp5}
\end{figure}

The spectral analysis was performed with XSPEC, version 12.5 \citep{Arnaud1996}
with the Band GRB function \citep{Band1993}:
$f(E) \propto E^{\alpha}\exp(-(2+\alpha)E/E_{\mathrm{peak}})$ for
$E < E_{\mathrm{peak}}(\alpha-\beta)/(2+\alpha)$, and $f(E) \propto E^{\beta}$ for
$E \geq E_{\mathrm{peak}}(\alpha-\beta)/(2+\alpha)$ where
$\alpha$ is the power-law photon index, $E_{\mathrm{peak}}$
is the peak energy in the $\nu F_\nu$ spectrum, and $\beta$ is the photon index
at higher energies. The spectral model was normalized to the energy flux
in the 20~keV--10~MeV range, a standard band for the KW GRB spectral analysis.

\begin{figure}[t!]
\centering
\includegraphics[width=0.5\textwidth]{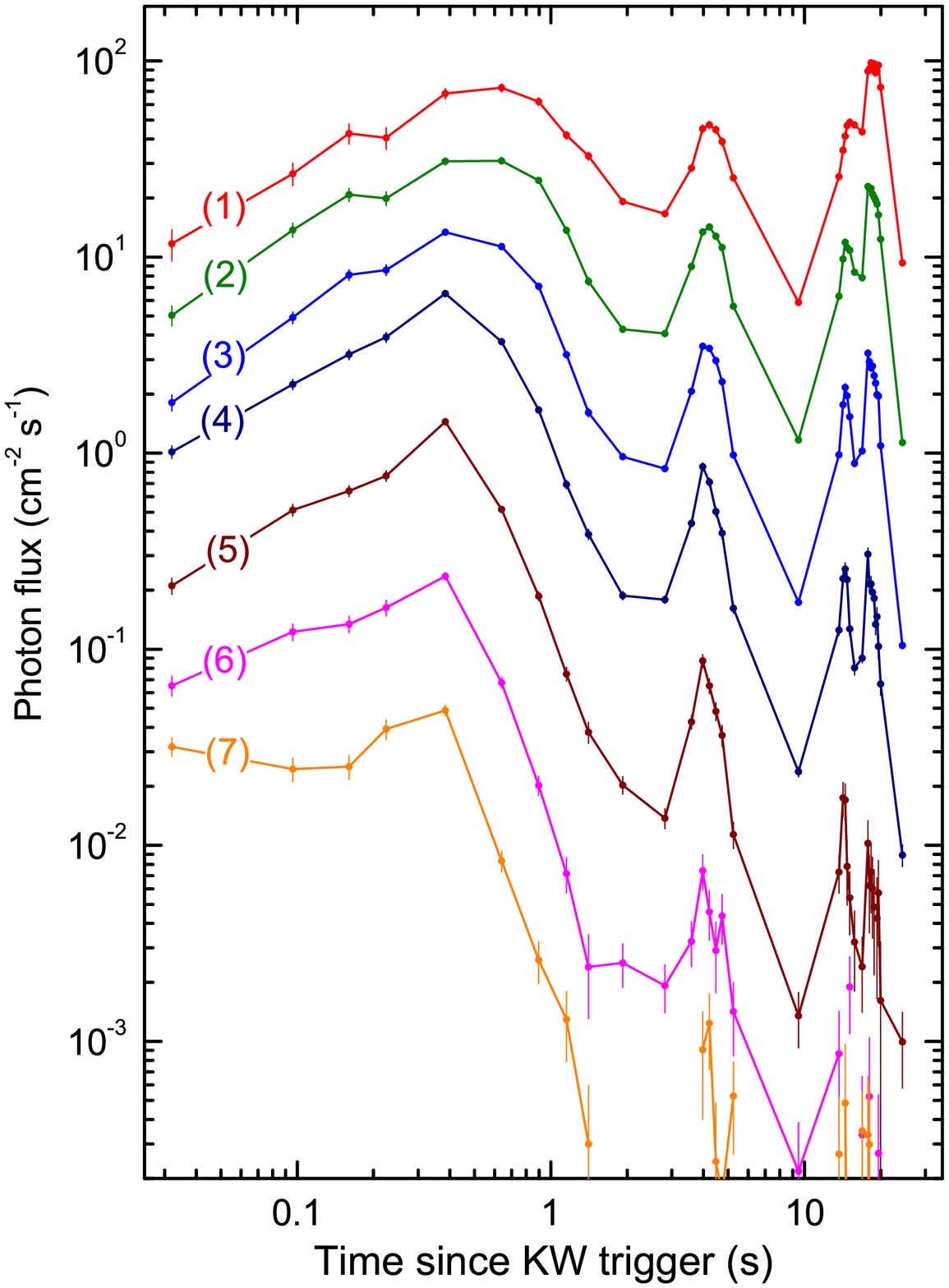}
\caption{Evolution of the photon flux during the prompt phase of \GRB.
The flux is calculated from the KW time-resolved spectral fits
and plotted vs. time for the following energy ranges:
20--50~keV (plot 1), 50--100~keV (2), 100--200~keV (3),
200--500~keV (4), 500--1000~keV (5), 1--2~MeV (6),
and 2--18~MeV (7). The plots, except (1), are consecutively
scaled down by half an order of magnitude for viewing convenience.\\
}
\label{FigPhotons}
\end{figure}

\begin{figure}
\centering
\includegraphics[width=0.5\textwidth]{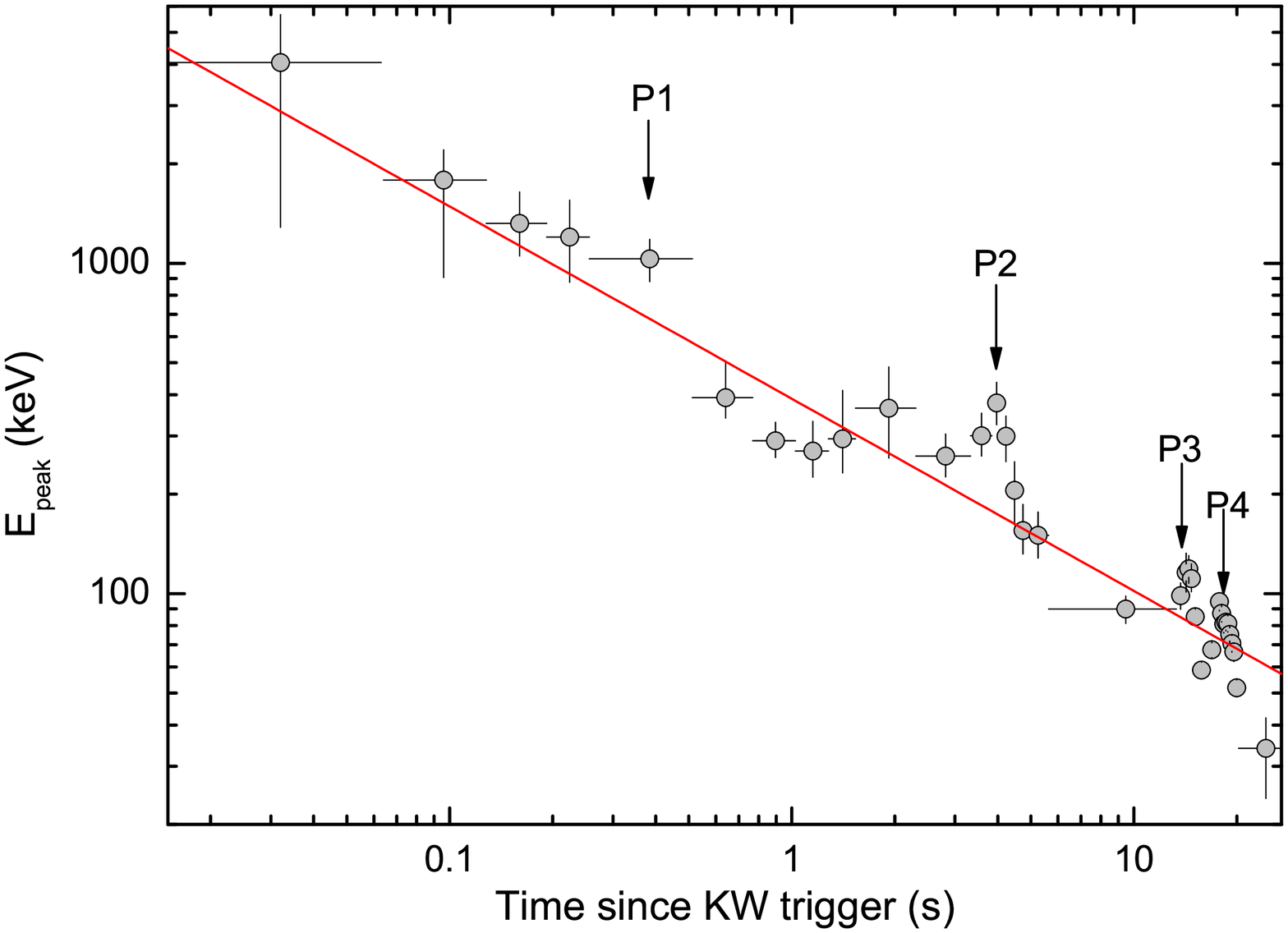}
\caption{General trend of $E_{\mathrm{peak}}$ evolution from the KW time-resolved spectral analysis
(circles). Spectra at four major emission peaks in the KW light curve are marked:
P1 (spectrum~5), P2 (spectrum~13), P3 (spectrum~21),
and P4 (spectrum~26). The best power-law approximation from $T_0$ to $T_0+20$~s
is shown with the solid line ($E_{\mathrm{peak}}(t)\propto t^{-0.6}$).\\
}
\label{FigEpeak}
\end{figure}

Typically, the raw count rate spectra were rebinned to have at least
20 counts per energy bin to ensure Gaussian-distributed errors
and the correctness of the $\chi^2$ statistic.
Spectra 3, 4, and 5 have good count statistics in the MeV band,
which allowed us to study the hard emission with a minimal channel binning.
For these spectra, the \texttt{cstat} and \texttt{pgstat} options of XSPEC were also used
and we found the fit results to be consistent with those obtained by the first method.
At the very high count rate observed in the \GRB\, initial pulse,
the differential nonlinearity (DNL) of the instrument's analog-to-digital
converters must be taken into account as a source of systematic errors in the count spectra.
To account for the known level of uncertainty in the DNL,
we added up to 25\% systematics to statistical errors for 256-ms spectra covering the initial pulse.
Also, channels below 50~keV were excluded from fits for spectrum~5 since the influence
of a pulse-pileup effect on the low-energy part of this spectrum is not negligible.
The same precautions were taken for time-averaged spectra, which include the initial pulse.

Results of the KW time-resolved spectral analysis are listed in Table~\ref{TableKWSpecRes}.
A good quality of fit is achieved for the majority of the spectra, which enables us to construct the temporal behavior
of the model parameters ($\alpha$, $\beta$, $E_{\mathrm{peak}}$) and to trace
in detail the evolution of the spectral composition of radiation
over the course of the burst (Figure~\ref{FigKWspRes}).

Spectra 1--4 were measured at the onset of the extremely intense initial pulse
(Figure~\ref{FigKWlc2}). The emission at this moment is very hard;
$E_{\mathrm{peak}}$ reaches the highest value for the burst ($\sim4$~MeV)
in the first 64-ms interval after the trigger.
As the intensity surges up, $E_{\mathrm{peak}}$ starts to decrease gradually,
but the flattening low-energy photon index, $\alpha$, indicates a spectrum
enriched by higher-energy photons.
Spectrum 5 describes the culmination of the pulse ($T_0+0.256$--$T_0+0.512$~s).
At this time, $E_{\mathrm{peak}}$ remains above 1~MeV and the energy flux, averaged over 256~ms, reaches a huge value of
$\sim7.8\times10^{-4}$~erg~cm$^{-2}$~s$^{-1}$.
The falling edge of the initial pulse (spectra 6--10, $T_0+0.512$--$T_0+2.304$~s)
is described by the further decrease in $E_{\mathrm{peak}}$ down to $\sim260$~keV and by
a slightly time-delayed steepening of $\alpha$ to $\sim-1.3$.

At the peak rate, no obvious high-energy cutoff was observed up to
the instrument's upper energy threshold (Figure~\ref{FigKWsp5}).
For the initial 64~ms spectra 1 and 2 taken individually, the high-energy photon
index, $\beta$, is poorly constrained. However, after applying a minimal time-averaging,
the fit to the average spectrum 1+2 yields $\beta=-2.35_{-0.79}^{+0.32}$,
consistent with the indices obtained for the subsequent spectra 3, 4, 5 and 6.
Thus, no significant spectral variation is observed above several
MeV during the onset and the culmination of the initial pulse.
This result is illustrated by plot~7 in Figure~\ref{FigPhotons},
where photon light curves for different energy ranges are shown as
constructed from the unfolded spectra. From this figure, one can see
that the MeV emission decays dramatically with the decay of the initial pulse
and has almost ceased after $\sim T_0+5$~s.

As already mentioned when discussing the KW light curves
and their hardness ratio behavior, a correlation between the radiation intensity and
its hardness is notable for the whole history of the burst.
Onsets of all four major emission pulses are characterized by an apparent rise in $E_{\mathrm{peak}}$
and by a pronounced flattening of the low-energy part of the spectrum, with an opposite
pattern in the decaying emission phases.
At the same time, there is a clear trend of spectral softening from pulse to pulse:
the top value of $E_{\mathrm{peak}}$ reached during each of the four major emission pulses
steadily declines in time from $\approx$4~Mev for the initial pulse to $\approx$380~keV, $\approx$130~keV,
and $\approx$105~keV for pulses $P_2$, $P_3$, and $P_4$, respectively.
The general trend of the peak energy evolution in the prompt emission phase
can be roughly described by a power-law slope $E_{\mathrm{peak}}(t)\propto t^{-0.6}$ (Figure~\ref{FigEpeak}).

The behavior of the high-energy photon index, $\beta$, is consistent with the general
tendency of the emission softening: $\beta$ steepens, displaying weak variations, from $\approx-2.3$
at the onset of the burst to $\beta\leq-3$ soon after the onset of Phase~II and later.

\subsubsection{Time-averaged Spectra and Energetics in the Prompt Emission}\label{sec:tienerg}

We analyzed the time-averaged spectrum of the entire prompt phase of the emission
and its separate parts in the same way as described in the previous section.
Among the tested models, the Band function is the only model that adequately describes
the shape of the spectrum. The results of the fits are summarized in
Table~\ref{TableKWSpecInt}.

\begin{deluxetable*}{lcccccc}
\tablewidth{0pt}
\tablecaption{\KW\, Time-averaged Spectral Fits with the Band Function
\label{TableKWSpecInt}}
\tablehead{
\colhead{Interval} & \colhead{$\alpha$} & \colhead{$\beta$} & \colhead{$E_{\mathrm{peak}}$} & \colhead{$\chi^2/$dof} & \colhead{Fluence\tablenotemark{a}} \\
\colhead{from $T_0$(s)} & \colhead{} & \colhead{} & \colhead{(keV)} & \colhead{} & \colhead{($10^{-4}$erg~cm$^{-2})$}
}
\startdata
0.000--28.416 & -1.64$_{-0.05}^{+0.06}$ & -2.25$_{-0.09}^{+0.09}$ & 340$_{-60}^{+70}$ & $77.5/81$ & 7.78$_{-0.45}^{+0.46}$ \\
\\
0.000--2.304 & -0.95$_{-0.05}^{+0.05}$ & -2.41$_{-0.12}^{+0.10}$ & 990$_{-90}^{+100}$ & $103/80$  & 4.03$_{-0.11}^{+0.11}$\\
(Initial pulse) &  &  &  &  &  \\
\\
0.000--13.312& -1.12$_{-0.08}^{+0.08}$ & -2.28$_{-0.10}^{+0.08}$ & 630$_{-100}^{+160}$ & $54.4/81$  & 6.09$_{-0.38}^{+0.34}$\\
(Phase~I) &  &  &  &  &  \\
\\
13.312--28.416 & -1.2$_{-0.1}^{+0.2}$ & -3.3$_{-0.2}^{+0.2}$ & 78$_{-3}^{+3}$ & $86.2/83$ & 1.57$_{-0.27}^{+0.33}$ \\
(Phase~II) &  &  &  &   &  \\
\enddata
\tablenotetext{a}{\footnotesize{In the 20~keV--10~MeV energy band.}}\\
\end{deluxetable*}

It should be emphasized that, having $E_{\mathrm{peak}}\approx340$~keV,
$\alpha\approx-1.6$, and $\beta\approx-2.3$, the overall time-integrated
spectrum ($T_0$ to $T_0+28.416$~s) is, indeed, the ``average'' one
and does not reflect the spectral composition of the emission at
any particular phase of the burst.
As expected from the time-resolved spectral analysis,
the average spectra of the first ($T_0\,-\,T_0+13.312$~s, Phase~I)
and the second ($T_0+13.312\,-\,T_0+28.416$~s, Phase~II) pairs
of overlapping pulses are essentially different.
With only the low-energy photon index being close, $\alpha\simeq-1.2$,
the peak energy differs between these phases of the burst almost
by an order of magnitude ($E_{\mathrm{peak}}\approx$630~keV and $\approx$78~keV, respectively).
Also, the high-energy photon index of the time-averaged spectrum
is substantially softer in Phase~II than in Phase~I
($\beta\approx-3.3$ and $\beta\approx-2.3$, respectively).
The average spectrum of the initial pulse ($T_0\,-\,T_0+2.304$~s)
is very hard: $E_{\mathrm{peak}}\approx$1~MeV, $\alpha\approx-1$, and $\beta\approx-2.4$.

Based on the results of the spectral and temporal analysis,
we calculated the time-integrated and peak energy flux
of the prompt $\gamma$-ray emission of \GRB.
In the 20~keV--10~MeV band, which is standard for
\KW, the total energy fluence, $S$,
from $T_0$ to $T_0+28.416$~s, is $(7.8\pm 0.5)\times10^{-4}$~erg~cm$^{-2}$.
More than half of this energy ($\sim55$\%) is released
in the initial pulse, $\sim80$\% during the hard Phase~I,
and only $\sim1/5$ of the total fluence comes from the much softer Phase~II.

The 64~ms peak energy flux, $F_{\mathrm{max}}=(9.2\pm0.4)\times 10^{-4}$~erg~cm$^{-2}$~s$^{-1}$,
is reached in the interval starting 0.384~s after the trigger,
at the culmination of the initial pulse (see Figure~\ref{FigKWlc2}).
As for the exceptionally high count rate,
the $F_{\mathrm{max}}$ is the highest among $>$2000 GRBs detected by \KW\, so far.
The 64-ms peak flux for Phase~II is found to be $\sim$20~times lower,
$(4.5\pm 0.5)\times 10^{-5}$~erg~cm$^{-2}$~s$^{-1}$,
in the interval starting at $T_0+17.536$~s,
at the peak of the last, softest pulse in the light curve.

\subsection{Extended Emission in $\gamma$-rays}
\label{sec:extended}

\begin{figure}
\centering
\includegraphics[width=0.5\textwidth]{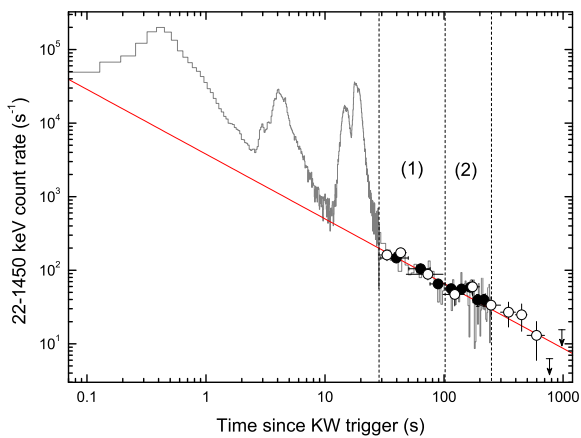}
\caption{Prompt and extended phases of the \GRB\, $\gamma$-ray emission (Section~\ref{sec:extended}).
Gray solid line: \KW\, count rate with 64~ms resolution before $T_0+28$~s and 2.944~s later).
Time-averaged KW and SPI-ACS data after $T_0+28$~s, are shown by filled and open circles, respectively;
the SPI-ACS counts are normalized to the KW counts in the 22--1450~keV energy range.
The straight solid line shows the best power-law fit to the extended emission light curve, $N(t)\propto t^{-0.88\pm0.05}$;
the `bump' around $T_0+165$~s has been discarded to derive the best fit value.
Time intervals (1) and (2) between the vertical dashed lines are used for the KW late-time spectral fits.\\
}
\label{FigKWext}
\end{figure}

It can be seen from the entire time history recorded by \KW\, (Figure~\ref{FigKWext})
that the extremely bright pulsed phase of \GRB\, ends $\sim$30~s after the trigger.
However, a stable excess in the count rate over the background level
was detected by the instrument until the end of the measurements at $T_0+250$~s.

At this final phase of the $\gamma$-ray emission, the light curve
shows a smooth temporal decay with a photon flux several orders
of magnitude fainter and a substantially changed spectrum shape, as compared to the prompt phase.
The ``transitional'' spectrum~35 (from $T_0+20.224$~s to $T_0+28.416$~s)
is the last one which requires a ``curved'' spectral model in the fits.
The subsequent time-resolved spectra were measured by KW out to $T_0+102.144$~s.
Since the emission is weak, the individual spectra have poor signal-to-noise ratio,
but may be adequately described by a simple power-law (PL)
$f(E) \propto E^{-\Gamma}$, with a photon index, $\Gamma$, of 1.7--2.3.
The average spectrum from $T_0+28.416$~s to $T_0+102.144$~s (interval 1 in Figure~\ref{FigKWext})
extends to $\sim$9~MeV and is best fit by a simple power law with $\Gamma=2.00\pm0.12$
(1$\sigma$ errors hereafter in this section) and $\chi2=106/99$~d.o.f.
The final portion of the KW data for \GRB\, is available from the background mode
light curves, which are measured out to $T_0+250$~s (interval 2 in the same figure).
The time histories recorded in the G1, G2, and G3 energy bands are, in effect,
a continuous three-channel spectrum covering the 22--1450~keV energy range.
The power-law fit to the average three-channel spectrum from $T_0+100$~s to $T_0+250$~s
yields $\Gamma=2.18\pm0.24$ ($\chi^2=0.40/1$~d.o.f.), suggesting that there is
no spectral evolution from the preceding interval 1, for which the same method gives
$\Gamma=1.96\pm0.13$ ($\chi^2=0.96/1$~d.o.f.); this is consistent with the index obtained
for the multi-channel spectrum.

In the {\it INTEGRAL} SPI-ACS data ($\sim$80~keV--10~MeV energy band),
the $\gamma$-ray emission of \GRB\, can be traced to $\sim~T_0+700$~s,
after which the source flux becomes indistinguishable from the unstable background.
Although the energy responses of the KW and SPI-ACS detectors are quite different,
the stable power-law shape of the spectrum allowed us to extend the late-time
$\gamma$-ray light curve beyond the end of the KW observation (open symbols in Figure~\ref{FigKWext}).
To ensure the correctness of the extrapolation, we normalized the count rate between
the two instruments using counts accumulated in the interval from $T_0+50$ to $T_0+250$~s,
well beyond the cessation of the prompt emission.
For this interval the KW three-channel spectral fit yields $\Gamma=2.00\pm0.16$ ($\chi^2=0.51/1$~d.o.f.);
in the following analysis we assume this slope and the corresponding spectral index,
$\beta_{\gamma}\equiv\Gamma-1=1.00\pm0.16$, for the extended $\gamma$-ray emission.

A temporal power-law (PL) fit to the combined light curve, $N(t)\propto t^{-\alpha}$,
yields, in the $T_0+30$~s--$T_0+700$~s time interval,
the decay index $\alpha_{\gamma}=0.84\pm0.05$ ($\chi^2=19.1/15$~dof).
A count rate excess over this slope is found around $T_0+165$~s.
Since the ``bump'' is present at the $\sim2\sigma$ level in both KW and SPI-ACS data,
it is unlikely to be a simple count rate fluctuation.
Excluding the corresponding KW and SPI-ACS data points from the fit
results in a better $\chi^2=10.8/13$~d.o.f., while the temporal index changes
only marginally to $\alpha_{\gamma}=0.88\pm 0.05$ (the solid line in Figure~\ref{FigKWext}).

Traced back in time, this slope lies just below the count rate observed
at the KW light curve minimum around $T_0+11$~s, which may suggest
that the power-law component emerges before the rise of Phase~II of the prompt emission.
A power-law fit to a three-channel spectrum constructed
for the narrow time interval from $T_0+9.728$~s to $T_0+11.776$~s
yields $\Gamma=2.01\pm0.03$ ($\chi^2=0.27/1$~d.o.f.), thus
favoring this hypothesis.

Assuming a power-law spectrum with $\Gamma$=2,
we estimate the 22--1450~keV energy fluence of the \GRB\, extended emission
from $T_0+28.416$~s to $T_0+700$~s to be $(2.6\pm0.5)\times 10^{-5}$~erg~cm$^{-2}$,
or $\approx0.3$\% of the energy in the prompt phase of the burst.
\\
\subsection{\SW\, Afterglow Observations}
\label{sec:afterglow}

\subsubsection{\SW/XRT Observations}

\begin{deluxetable*}{cccccc}
  \tablewidth{0pc} 	      	
  \tabletypesize{\scriptsize}
  \tablecaption{\SW/XRT observations of \GRB\, \label{TableXRTobs}}
  \tablehead{	
  \colhead{Spectrum} & \colhead{Sequence}  &    \colhead{Start Time  (UT)} &    \colhead{End Time   (UT)} &
  \colhead{Exposure}  \\
  \colhead{} & \colhead{} &  \colhead{(yyyy-mm-dd hh:mm:ss)} & \colhead{(yyyy-mm-dd hh:mm:ss)} &
  \colhead{(s)} \\
\colhead{(1)} 	 & \colhead{(2)} 	 & \colhead{(3)} 	 & \colhead{(4)} 	 & \colhead{(5)}
}
  \startdata	
$-$&00020186001\tablenotemark{a}    &	2011-09-19 20:30:55	&	2011-09-19 20:59:05	&	1675	\\
 \hline
1&00020186002\tablenotemark{b}	    &	2011-09-20 03:16:34	&	2011-09-20 07:39:25	&	2598	\\
 \hline
2&00020187001		&	2011-09-20 16:01:31	&	2011-09-20 20:59:48	&	6925	\\
 \hline
3&00020187002		&	2011-09-21 00:03:19	&	2011-09-21 22:30:56	&	8342	\\
 \hline
4&00020187003		&	2011-09-22 01:45:13	&	2011-09-22 21:13:56	&	8191	\\
 \hline
5&00020187004		&	2011-09-23 00:12:46	&	2011-09-23 00:26:56	&	830     \\
&00020187005		&	2011-09-23 06:36:46	&	2011-09-23 21:02:57	&	4671	\\
&00020187006		&	2011-09-24 01:57:11	&	2011-09-24 05:20:57	&	2023	\\
 \hline
6&00020187007		&	2011-09-24 06:26:48	&	2011-09-24 21:23:58	&	6928	\\
 \hline
7&00020187008		&	2011-09-25 00:28:30	&	2011-09-25 19:39:57	&	6807	\\
&00020187009		&	2011-09-26 00:32:23	&	2011-09-26 23:10:57	&	8863	\\
 \hline
8&00020187010		&	2011-09-27 00:09:44	&	2011-09-27 23:06:57	&	7271	\\
&00020187011		&	2011-09-28 00:38:07	&	2011-09-28 18:30:56	&	7702	\\
&00020187012		&	2011-09-29 00:46:15	&	2011-09-29 08:29:58	&	4566	\\
 \hline
9&00020187013		&	2011-09-30 00:22:28	&	2011-09-30 18:37:57	&	4450	\\
&00020187014		&	2011-10-01 00:25:45	&	2011-10-01 15:01:55	&	5050	\\
&00020187015		&	2011-10-02 03:43:33	&	2011-10-02 10:24:56	&	4864	\\
&00020187016		&	2011-10-03 07:00:51	&	2011-10-03 12:04:56	&	4929	\\
&00020187017		&	2011-10-04 00:59:57	&	2011-10-04 07:21:56	&	4518	\\
&00020187018		&	2011-10-05 00:44:43	&	2011-10-05 18:59:56	&	4185	\\
&00020187019		&	2011-10-06 02:24:07	&	2011-10-06 23:45:56	&	3673	\\
&00020187020		&	2011-10-07 15:31:38	&	2011-10-07 22:19:56	&	4964	\\
&00020187021		&	2011-10-08 12:11:36	&	2011-10-08 17:27:56	&	4899	\\
&00020187022		&	2011-10-09 10:39:43	&	2011-10-09 15:31:18	&	414     \\
&00020187023		&	2011-10-10 09:07:53	&	2011-10-10 13:00:57	&	5092	\\
&00020187024		&	2011-10-11 09:33:12	&	2011-10-11 17:30:57	&	4884	\\
&00020187025		&	2011-10-12 06:04:30	&	2011-10-12 22:47:56	&	4919	\\
&00020187026		&	2011-10-13 00:07:46	&	2011-10-13 21:09:56	&	4786	\\
&00020187027		&	2011-10-14 12:37:29	&	2011-10-14 22:56:57	&	5115	\\
&00020187028		&	2011-10-15 07:54:36	&	2011-10-15 19:35:56	&	5208	\\
&00020187029		&	2011-10-16 00:21:58	&	2011-10-16 22:55:24	&	4190	\\
&00020187030		&	2011-10-17 11:29:47	&	2011-10-17 18:07:58	&	3565	\\
&00020187031		&	2011-10-18 08:06:28	&	2011-10-18 18:12:56	&	5072	\\
&00020187032		&	2011-10-19 14:50:41	&	2011-10-19 21:26:57	&	5466	\\
&00020187033		&	2011-10-20 18:13:31	&	2011-10-20 23:19:56	&	4069	\\
&00020187034		&	2011-10-21 10:22:09	&	2011-10-21 20:14:56	&	5040	\\
&00020187035		&	2011-10-22 16:42:29	&	2011-10-22 21:37:05	&	3377	\\
&00020187036		&	2011-10-23 07:11:17	&	2011-10-23 08:51:00	&	364     \\
&00020187037		&	2011-10-24 13:41:17	&	2011-10-24 18:32:40	&	4064	\\
&00020187038		&	2011-10-25 01:00:38	&	2011-10-25 07:42:57	&	5320	\\
&00020187039		&	2011-10-26 07:31:17	&	2011-10-26 21:59:57	&	4553	\\
&00020187040		&	2011-10-27 09:00:17	&	2011-10-27 19:04:57	&	4992	\\
&00020187041		&	2011-10-28 13:53:20	&	2011-10-28 19:09:58	&	4904	\\
&00020187042		&	2011-10-29 18:48:19	&	2011-10-29 19:13:56	&	1534	\\
&00020187043		&	2011-10-30 10:51:12	&	2011-10-30 19:17:58	&	8031	\\
&00020187044		&	2011-10-31 11:11:01	&	2011-10-31 20:56:56	&	8234	\\
&00020187045		&	2011-11-01 09:21:36	&	2011-11-01 17:48:49	&	8079	\\
&00020187046		&	2011-11-02 07:59:37	&	2011-11-02 22:44:58	&	8550	\\
&00020187047		&	2011-11-03 07:46:52	&	2011-11-03 21:09:56	&	7655	\\
&00020187048		&	2011-11-04 06:24:06	&	2011-11-04 18:02:57	&	8096	\\
&00020187049		&	2011-11-04 16:05:13	&	2011-11-05 22:58:56	&	9796	\\
&00020187050		&	2011-11-06 08:05:06	&	2011-11-06 23:03:56	&	9904	\\
&00020187051		&	2011-11-07 08:14:23	&	2011-11-07 23:04:18	&	8695	\\
 \hline
$-$&00020187052		&	2012-10-04 07:44:32	&	2012-10-04 23:59:54	&	2675 	\\
$-$&00020187053		&	2012-10-09 11:43:27	&	2012-10-09 11:46:55	&	188  \\
$-$&00020187054		&	2012-10-10 16:02:44	&	2012-10-10 22:48:55	&	3049 	\\
$-$&00020187055		&	2012-10-11 13:07:42	&	2012-10-11 23:07:55	&	5348 	\\
$-$&00020187056		&	2012-10-12 21:04:55	&	2012-10-13 19:59:54	&	9232 	\\
\enddata
    \tablenotetext{a}{\GRB\, was out of the XRT FoV.}
    \tablenotetext{b}{\GRB\, was detected close to the edge of the XRT FoV only during the initial two snapshots, for a total exposure of 1481 s.}
    \tablecomments{
    Column 1 lists the reference number of the spectrum extracted after merging
    the data of the observations within horizontal separators;
    sequences not used for spectral analysis are marked with a `$-$'.
    Column 2 lists the sequence numbers of all the \SW/XRT pointed observations of \GRB.
    Columns 3, 4 and 5 give information about start and stop time in UT and
    net exposure (after reduction and cleaning) of all observations,
    respectively.
   }
\end{deluxetable*}

The \GRB\, IPN error box was observed by the \SW\, X-ray Telescope (XRT; \citealt{Burrows2005})
with tiling strategy starting $\sim$83.0~ks after the KW trigger.
No X-ray point source was detected in the very first observation.
An unidentified X-ray source was detected close to the edge
of the XRT field of view (FOV) in the first two snapshots of
the second observation (starting $\sim$107.4~ks after the trigger)
and interpreted as the likely X-ray afterglow counterpart of \GRB\,
\citep{GCN12364}.
Subsequent pointed observations of this source revealed a power-law decay
of the flux and confirmed it as the counterpart \citep{GCN12376}.
The best XRT position of the source is the UVOT-enhanced position
(calculated using the XRT-UVOT alignment and by matching the UVOT field sources
to the USNO-B1 catalogue):
R.A.$(J2000)   = 32\fdg53869~(02^{h}~10^{m}~9.29^{s})$
decl.$(J2000)  = -27\fdg10576~(-27\arcdeg06\arcmin20.7\arcsec)$
with an uncertainty of 1.4\arcsec\, (radius, 90\% confidence).
The refined IPN ellipse (Section~\ref{sec:ipn}) encompasses the counterpart.

Table~\ref{TableXRTobs} reports the log of the \SW/XRT observations used for this work.
The source was observed by \SW/XRT in full frame Photon Counting mode
('pcw3') for a total of 48 days and
a total on-source exposure of $\sim~280$~ks in 2011,
and 9 days with on-source exposure of $\sim~20.5$~ks one year later
starting on October 4th in 2012 (see Section \ref{sec:uvot}).

\subsubsection{XRT Light Curve}
\label{sec:xrtlc}

\begin{figure}
\includegraphics[width=0.5\textwidth]{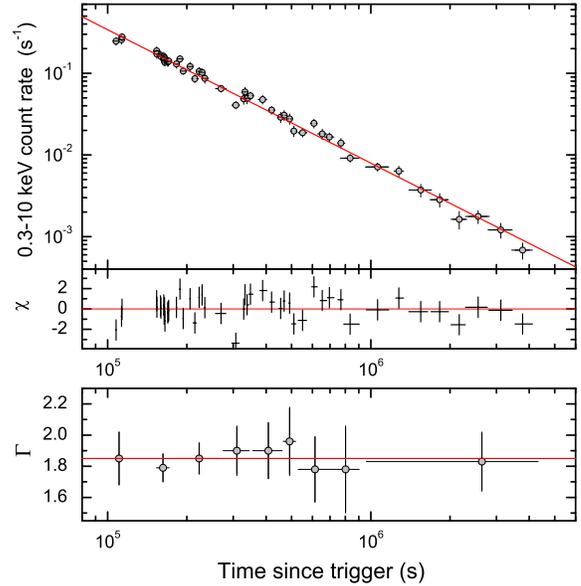}
\caption{Top panel: \SW/XRT count rate light curve (symbols) with the best power-law fit
(solid line) and residuals.
Bottom panel: temporal evolution of the photon index, $\Gamma$, of the time resolved
\SW/XRT spectra (symbols). The photon index of the time-integrated spectrum
($\Gamma=1.85$) is shown for reference with the solid line.\\
\label{FigXRTrate}}
\end{figure}

The \SW/XRT count rate light curve in the 0.3$-$10~keV band
has been built according to the light curve creation procedure
described in \cite{Evans2007}, requiring a minimum of 70 counts
per bin and dynamic binning. In the 2012 observations,
the source was not detected and only a 3$\sigma$ upper
limit of $1.28\times10^{-3}$~counts~s$^{-1}$ has been calculated
using the Bayesian method of \cite{Kraft1991}.
The best-fit model is a single power law with decay slope
$\alpha_{\mathrm{X}}=-1.63\pm0.02$ and $\chi^2=46.4/45$~d.o.f.
The count-rate light curve, together with best-fit model and
residuals, is shown in Figure~\ref{FigXRTrate}.

We tested the XRT afterglow light curve for the presence of a
temporal break by fitting the data with a broken power law (BPL) function:
\begin{equation}\label{eq:bpl}
  N(t)\propto
  \begin{cases}
  t^{-\alpha_1} &, (t<t_{b}) \\
  \\
  t_{b}^{(\alpha_2-\alpha_1)}\,t^{-\alpha_2} &, (t>t_{b}) \\
  \end{cases}
\end{equation}
where $t_b$ is the time of break in the light curve; $\alpha_1$ and $\alpha_2$ are the temporal indices
before and after the break, respectively.
An $F$-test shows that there is little evidence of an improvement
in the fit if a break is added ($F=4.62$, giving a null-hypothesis probability of 0.015).
This result has been checked using different binning criteria.
We generally find an F-test probability of $>10^{-2}$
suggesting, that a BPL model is not needed, with
a high significance, to fit the light curve.

We also tested a smoothly broken power law \citep{GS2002};
it results in an ambiguous fit and does not constrain the break.

\subsubsection{XRT Spectra}
\label{sec:xrtspec}

%
\begin{figure}
\includegraphics[width=0.5\textwidth]{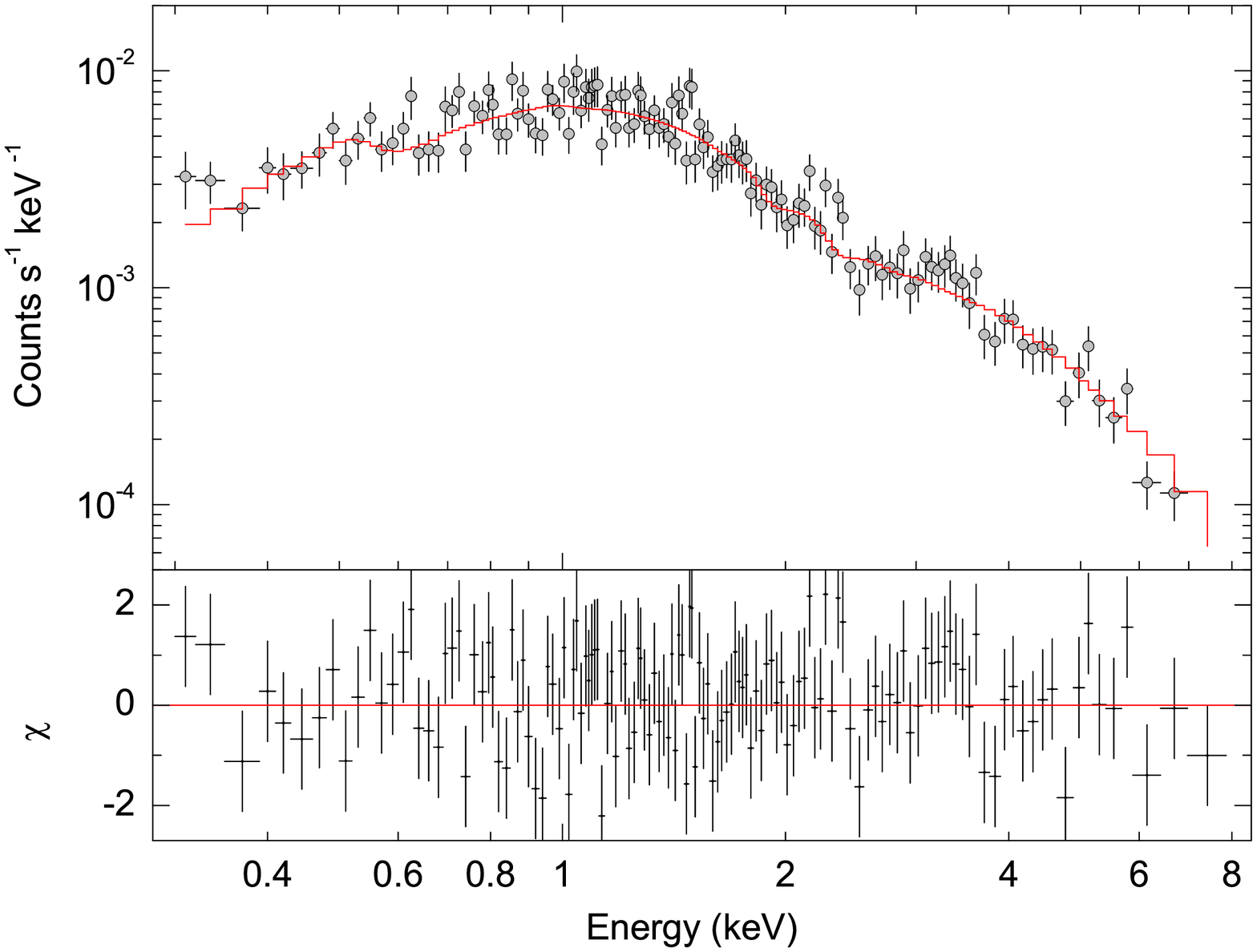}
\caption{Average spectrum of the 2011 \SW/XRT observations together
with the best-fit power-law model including Galactic and intrinsic absorption at the source and residuals (Section~\ref{sec:xrtspec}).
\label{FigXRTspec_int}}
\end{figure}

We extracted average spectra of the source and the background
after merging all the 2011 observations
(sequences from 00020186002 to 00020187051 in Table \ref{TableXRTobs},
covering times from $\sim$107.3~ks to $\sim$4.325~Ms after the trigger).
For the source extraction region, we used the 15-pixel-radius circular region
needed to remain within the XRT FOV in sequence 00020186002.

To perform time-resolved spectral analysis, we also extracted a set of nine
source and background spectra with comparable statistics. The
observations were grouped as shown in Table \ref{TableXRTobs}. The small source
extraction region with 15 pixel radius was used for spectrum 1
(coming from the highly offset initial observation)
and for spectrum 9, corresponding to the faintest state of the source.
The other spectra have been extracted from a circular 30-pixel-radius region.
All background spectra were extracted from the same region used for the
average background spectrum.

The spectra were all fit with an absorbed power law model with two
absorption components: Galactic absorption in the direction of the source
fixed at the value of 0.168$\times$\NHunits \citep{Kalberla2005}
and intrinsic absorption at the measured redshift of z=0.984.
A solar metallicity is assumed for both absorption components.

\begin{deluxetable}{ccccc}
\tablewidth{0pt}
\tablecaption{Absorbed Power-law Fits to the Time-resolved \SW/XRT Spectra
\label{TableXRTspec_par}}
\tablehead{
\colhead{Mean Epoch}      & \colhead{Photon Index} & \colhead{Observed Flux} & \colhead{$\chi^2$/d.o.f.}\\
\colhead{(ks from $T_0$)} & \colhead{$\Gamma$}     & \colhead{($10^{-13}$erg~cm$^{-2}$~s$^{-1}$)} & \colhead{}
}
\startdata
111 &   1.85$^{+0.17}_{-0.17}$ & $90.1^{+15.9}_{-12.4}$ & 7.64/9\\
162	&   1.79$^{+0.09}_{-0.09}$ & $60.5^{+5.0}_{-4.3}$   & 36.7/41\\
223	&   1.85$^{+0.10}_{-0.10}$ & $40.8^{+3.7}_{-2.9}$   & 33.5/34\\
310	&   1.86$^{+0.19}_{-0.18}$ & $20.0^{+2.7}_{-2.4}$   & 21.6/17\\
408	&   1.89$^{+0.20}_{-0.19}$ & $14.5^{+2.6}_{-1.9}$   & 7.37/11\\
491	&   1.96$^{+0.21}_{-0.22}$ & $10.3^{+2.1}_{-1.5}$   & 8.56/8\\
613	&   1.78$^{+0.21}_{-0.21}$ & $7.9^{+1.6}_{-1.2}$    & 11.9/12\\
802	&   1.78$^{+0.27}_{-0.28}$ & $4.8^{+1.2}_{-0.89}$   & 12.7/10\\
2643&   1.83$^{+0.18}_{-0.19}$ & $0.78^{+0.13}_{-0.10}$ & 16.7/18\\
\enddata

\tablecomments{
        The flux is given in the (0.3--10)~keV range.
        Galactic and intrinsic $N_\mathrm{H}$ at redshift z=0.984 have been
        fixed to 0.168$\times$\NHunits and 2.33$\times$\NHunits, respectively.
}
\end{deluxetable}

The XRT average spectrum is best fit by an intrinsic absorption
$N_{\mathrm{H,i}}=2.33_{-0.53}^{+0.58}\times$\NHunits,
a photon index $\Gamma$ = 1.85$_{-0.07}^{+0.07}$ and
$\chi^2_r$ = 1.02 (134 d.o.f.).
The average [observed] unabsorbed flux in the 0.3$-$10 keV band
is [5.3] 6.2$\times 10^{-13}$\flux.
In Figure~\ref{FigXRTspec_int}, we show the data and best fit model.
From this fit we derived a rate to [observed] unabsorbed flux conversion
factor in the 0.3$-$10 keV band of [4.58] 5.15 $\times 10^{-11}$ erg cm$^{-2}$.

The time-resolved spectra are all best fit with the same absorbed power law model.
Fits performed with a free $N_{\mathrm{H,i}}$ parameter always lead to best-fit values
consistent within errors with the ones previously obtained for the average spectrum.
We then fixed $N_{\mathrm{H,i}}$ to 2.33$\times$\NHunits.
Final results for the photon index, average observed flux, and $\chi^2_r$ for
the time-resolved spectra are listed in Table~\ref{TableXRTspec_par};
the index is shown in the lower panel of Figure~\ref{FigXRTrate},
where we see no evidence of spectral evolution.

\subsubsection{Swift-UVOT Observation and Analysis}
\label{sec:uvot}

%
\begin{figure}
\centering
\includegraphics[width=0.5\textwidth]{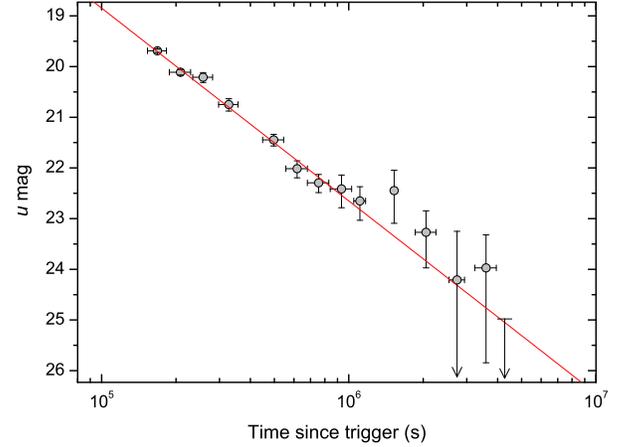}
\caption{UVOT light curve of \GRB\, afterglow (Table~\ref{TableUVOT}).
The underlying galaxy flux is subtracted and the data are normalized
to the count rate in the $u$ filter.
The solid line shows a best power-law fit with a slope of $1.52\pm0.05$.
}
\label{FigUVOT}
\end{figure}

\begin{deluxetable}{cccc}
\tablewidth{0pt}
\tablecaption{UVOT Combined Normalized Photometry
\label{TableUVOT}}
\tablehead{
\colhead{$T_{mid}$} & \colhead{Half-exposure} & \colhead{Count Rate} & $u$\\
\colhead{(ks from $T_0$)} & \colhead{(ks)} & \colhead{(s$^{-1}$)} & \colhead{(mag)}
}
\startdata
168.1	& 14.8	& 0.2895$\pm$0.0099\tablenotemark{a}& 19.69$^{+0.04}_{-0.04}$\\
208.5	& 20.7	& 0.1961$\pm$0.0098	& 20.11$^{+0.06}_{-0.05}$\\
257.5	& 23.7	& 0.1788$\pm$0.0160	& 20.21$^{+0.10}_{-0.09}$\\
326.8	& 29.6	& 0.1090$\pm$0.0122	& 20.75$^{+0.13}_{-0.11}$\\
496.7	& 48.5	& 0.0574$\pm$0.0061	& 21.44$^{+0.12}_{-0.11}$\\
618.2	& 61.8	& 0.0340$\pm$0.0052	& 22.01$^{+0.18}_{-0.16}$\\
755.0	& 74.9	& 0.0263$\pm$0.0043	& 22.29$^{+0.20}_{-0.17}$\\
933.8	& 92.8	& 0.0235$\pm$0.0068	& 22.41$^{+0.37}_{-0.28}$\\
1108.7	& 61.2	& 0.0189$\pm$0.0056	& 22.65$^{+0.38}_{-0.28}$\\
1525.1	& 38.5	& 0.0228$\pm$0.0102	& 22.45$^{+0.64}_{-0.40}$\\
2054.1	& 197.7	& 0.0107$\pm$0.0051	& 23.27$^{+0.70}_{-0.42}$\\
2741.1	& 197.1	& 0.0045$\pm$0.0064	& 24.20$^{+nan}_{-0.98}$\\
3591.8	& 358.7	& 0.0056$\pm$0.0046	& 23.96$^{+1.81}_{-0.65}$\\
4143.1	& 182.7	& $<$~0.0022         & $>$~24.98\\
33366.3	& 410.9	& $<$~0.0034         & $>$~24.51\\
\enddata
\tablenotetext{a}{\footnotesize{1$\sigma$ errors in this table.}}\\
\end{deluxetable}

{\it Swift's} Ultraviolet Optical Telescope (UVOT; \citealt{2000SPIE.4140...76R, 2004SPIE.5165...262R, 2005SSRv..120...95R})
began observing the burst approximately 153~ks after the KW trigger.
Data were initially taken in UVOT's $white$ filter, but late-time data (200-500~ks after the trigger)
were also taken in UVOT's four ultraviolet filters ($u$, $uvw1$, $uvm2$ and $uvw2$) until \SW\, ceased observations.

Photometry was measured with UVOTSOURCE using the calibrations of \cite{2008MNRAS.383..627P},
\cite{Breeveld2010}, and \cite{2011AIPC.1358..373B}. A combined light curve from all detections,
normalized to the count rate in the $white$ filter, suggests a late-time flattening of the light curve. In
order to exclude contamination from nearby sources or the possible host galaxy, we requested additional
observations with {\it Swift}/UVOT. Since the best coverage of the afterglow was in the $white$ and
$u$ filters, we requested 10~ks of observations for each of these two filters.
We examined these data for the presence of a host galaxy and, within the source aperture, we found a
candidate object. Analysis of this object provided a 5.3$\sigma$ detection in $white$ with a magnitude
of $22.63\pm0.21$ and a 2.6$\sigma$ detection in the $u$ band at a magnitude of $22.49\pm0.42$
(1$\sigma$ errors).
These findings are in agreement with E13, where results of detailed studies of the host can be found.

Using the late-time UVOT observations, we subtracted the underlying flux from the $white$ and $u$-band data.
Figure~\ref{FigUVOT} and Table~\ref{TableUVOT} show the combined light curve from these two bands, normalized to the $u$ filter and co-added.
The host-subtracted light curve shows a power-law decay with a slope $\alpha_{\mathrm{opt}}=-1.52\pm0.09$ ($\chi^2=9.35/11$~d.o.f)
out to 3.4~Ms after the trigger. Introducing a temporal break does not improve the fit:
the best BPL fit yields $\chi^2=9.27/9$~d.o.f and a very high (96\%)
probability of chance improvement over the simple power law. The pre- and post-break indices
$\alpha_{1}=1.46\pm0.19$ and $\alpha_{2}=1.55\pm0.16$ are indistinguishable within
1$\sigma$ errors, and the break time, $t_{b}=258\pm415$~ks, cannot be determined.
A search for an achromatic break in the UV/optical and X-ray afterglow was performed
by fitting the UVOT and XRT data with the BPL model simultaneously.
We put minimal restrictions on the procedure and tied only the key model parameter,
$t_b$, between the optical and the X-ray light curves.
Nevertheless, no limits on the break were found in the fits and an F-test
probability $\approx 0.09$ suggests that the BPL law model is not needed.

\begin{figure}
\centering
\includegraphics[width=0.5\textwidth]{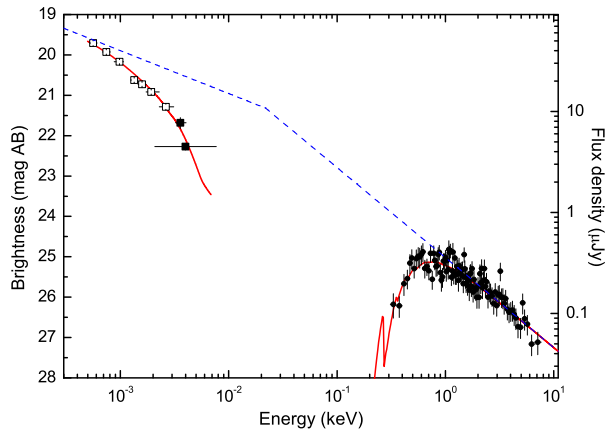}
\caption{Broadband SED of \GRB\, afterglow.
The SED is built for an epoch at 300~ks after the trigger (Section~\ref{sec:sed})
using the XRT spectra (solid circles), the UVOT photometry (solid squares),
and the GROND host-subtracted photometric data from E13 (open squares).
The best BPL fit for the SMC dust model is shown
(solid line) together with the model corrected for extinction
and absorption effects (dashed line). The break is located
between the UVOT and XRT bands and the spectral slope change, $\beta_{X}-\beta_{\mathrm{opt}}\approx0.47$,
is consistent with a cooling break of the standard fireball model.
}
\label{FigSED}
\end{figure}

\subsubsection{Broadband SED}
\label{sec:sed}

In a companion paper (E13), a broadband spectral energy distribution (SED) of the \GRB\, afterglow was studied,
as constructed from the optical/NIR GROND photometry at a midtime of 194~ks,
and the XRT data between 140~ks and 250~ks.
From the best simple PL fit ($\chi^2=85/73$~d.o.f), the authors derived a line-of-sight extinction $A_V=0.16\pm0.06$\footnote[1]{E13 reported errors at the 1$\sigma$ confidence level.}~mag,
an intrinsic column density $N_{\mathrm{H,i}}=1.56_{-0.46}^{+0.52}\times$\NHunits,
and a broadband spectral slope $\beta_{OX}=0.70\pm0.02$;
they also found that a BPL modeling of the SED does not improve
the fit ($\chi^2=83/71$~d.o.f).

Using the GROND host-subtracted photometry from E13, we analyzed a broadband SED built
following the method of \cite{Schady2010} from the GROND data in the $g'r'i'z'JHK_S$ bands, the XRT data,
and the host-subtracted UVOT photometry in the \emph{u} and \emph{white} filters.
The SED was constructed for an epoch centered at 300~ks utilizing
the observations from 107~ks to 450~ks after the trigger.
Fits were made in XSPEC with the absorbed power-law model and three dust extinction
curves from \cite{Pei1992}: the Milky Way (MW) with $R_V$=3.08, the Large Magellanic Cloud (LMC)
 with $R_V$=3.16, and the Small Magellanic Cloud (SMC) with $R_V$=2.93.
The Galactic reddening is fixed to 0.02~mag \citep{Schlegel1998}
and the Galactic column density is set to $0.168 \times 10^{21}$~cm$^{-2}$.
The best fit with a simple PL ($\chi^2=155.3/111$~d.o.f.) is achieved for the SMC dust profile
and closely reproduces the results of E13 at 194~ks: $A_V=0.22\pm0.02$~mag,
$N_{\mathrm{H,i}}=1.65_{-0.38}^{+0.42}\times$\NHunits, and $\beta_{OX}=0.69\pm0.01$.
This slope and the implied column density are in poor agreement
with the ranges obtained in our analysis of both time-resolved
and time-averaged XRT spectra; this suggests that
a single PL does not describe the broadband SED adequately.
As distinct from E13, a BPL model significantly improves our fit,
with $\chi^2=128.4/109$~d.o.f. and a chance improvement probability over the simple PL of $3.1\times10^{-5}$.
The best BPL fit (Figure~\ref{FigSED}) is found for the same SMC dust profile;
it yields $A_V=0.35\pm0.09$~mag, $N_{\mathrm{H,i}}=2.94_{-0.66}^{+0.73}\times$\NHunits,
the break energy $E_b=0.022_{-0.013}^{+0.046}$~keV,
$\beta_{\mathrm{opt}}=0.42\pm0.18$, and $\beta_{X}=0.89_{-0.07}^{+0.08}$. This result is in agreement with
the XRT spectral analysis and a spectral slope change $\beta_{X}-\beta_{\mathrm{opt}}\approx0.47$ suggests
a cooling break between the UVOT and XRT spectral bands; the derived range of $E_b$ favors of this hypothesis.

The XRT+UVOT+GROND SED centered at 194~ks (the E13 epoch) was also tested and we found
that both simple and broken PL fits result in a 1$\sigma$ agreement with those for the SED at 300~ks.
Thus, the significant improvement found for the BPL fits in this work
may result from a combination of at least two factors. First, with the UVOT observations added,
the shape of the optical part of the SEDs at both epochs is better understood.
Second, the overall signal-to-noise ratio in the SED at 300~ks is considerably higher,
since it was built from the $\sim3.5$-times longer time span of observations.

We note that $A_V\sim0.35$ derived from the BPL fit is higher than
the line-of-sight extinction resulting from a simple PL,
but still much smaller than the average $A_V$ of the host galaxy's starlight ($\sim0.90$~mag) reported by E13.
This inconsistency has been discussed in E13; the authors consider two explanations:
(1) a clumpy geometry of dust within the host and (2) that the progenitor had enough time
to destroy local dust with its UV emission. Bearing in mind the huge brightness of \GRB,
the latter option can be naturally extended by the possibility that the GRB itself 
partially destroyed dust grains along the line of sight (see, e.g., \citealt{WD2000,Fruchter2001}).

\section{DISCUSSION}
\subsection{\GRB\, in the Cosmological Rest Frame}
\label{sec:energy}

\begin{figure}[t!]
\centering
\includegraphics[width=0.5\textwidth]{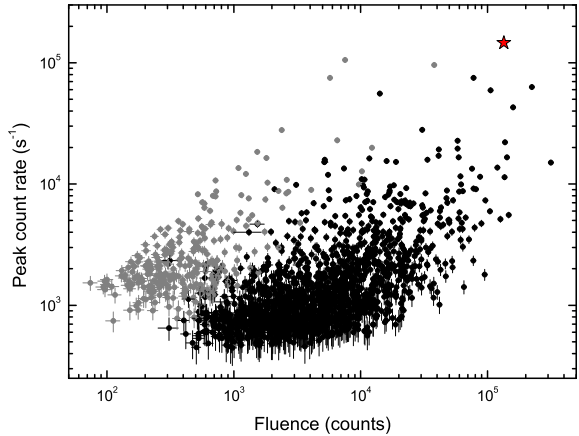}
\caption{Peak count rate vs. total recorded counts for 1834 \KW\, GRBs detected in 1994-2010
(Svinkin et al., in preparation): 1560 long (dark circles) and 274 short duration (gray circles);
\GRB\, is indicated by the star.
The counts are in the \KW\, G2+G3 energy band; the peak count rate is measured over 64~ms.
}
\label{FigKWcounts}
\end{figure}

\begin{figure*}
\centering
\includegraphics[width=0.9\textwidth]{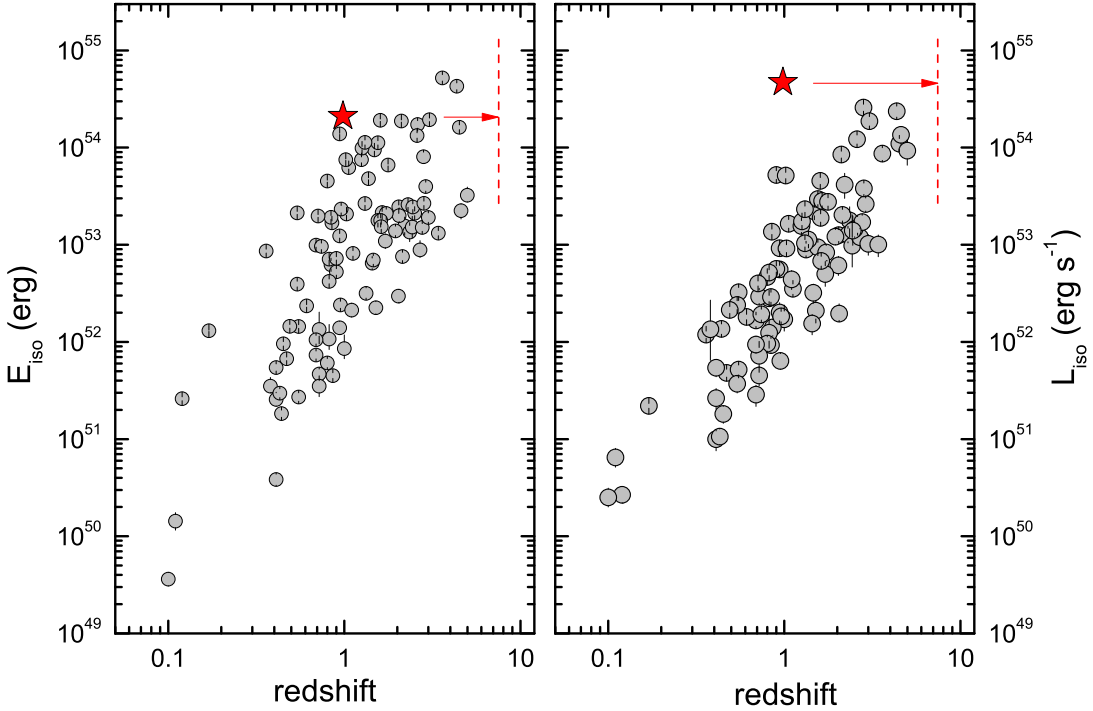}
\caption{
$E_{\mathrm{iso}}$ (left) and $L_{\mathrm{iso}}$ (right) of \GRB\, (stars) and \KW\, GRBs with known redshifts.
The gray circles show data for the KW bursts from Tsvetkova et al. (in preparation).
The dashed lines indicate the KW detection horizon for a \GRB-like burst (z$\sim$7.5).
}
\label{FigEisoLiso}
\end{figure*}

\KW\, started operation in 1994 November, and, in almost 19 years,
has detected more than 2000 $\gamma$-ray bursts, with virtually no bright GRBs having been missed.
Among these bursts, \GRB\, is, without a doubt, an outstandingly bright event (Figure~\ref{FigKWcounts}).
Its fluence, both in energy and count spaces, is among the half-dozen highest observed.
The burst's 64~ms peak count rate is unprecedented. Its peak energy flux is surpassed
only on the short 2~ms timescale by the two ultra-bright, short, hard bursts GRB~051103 and GRB~070201,
the candidate giant flares from soft $\gamma$-ray repeaters in the nearby galaxies
M81/82 and M31 \citep{Frederiks2008,Mazets2008,Hurley2010}.

The huge energy flux measured by KW suggests an enormous energy released in
the cosmological rest frame. Assuming a redshift of $z=0.984$ and a standard cosmology model
with $H_0=71$~km~s$^{-1}$~Mpc$^{-1}$, $\Omega_M=0.27$,
and $\Omega_{\Lambda}=0.73$, the luminosity distance $D_L$ is $2.0\times10^{28}$~cm.
Derived from the total energy fluence and the peak energy flux (Section~\ref{sec:tienerg}),
the isotropic-equivalent energy release in $\gamma$-rays, $E_{\mathrm{iso}}$, is $(2.1\pm0.1)\times10^{54}$~erg,
and the peak isotropic luminosity, $L_{\mathrm{iso}}$, is $(4.7\pm0.2)\times10^{54}$~erg~s$^{-1}$
(both quantities are in the rest-frame 1--10000~keV band).

These estimates make \GRB\, one of the most energetic and the most luminous $\gamma$-ray burst
observed since the beginning of the cosmological era in 1997.
Figure~\ref{FigEisoLiso} shows $E_{\mathrm{iso}}$ and $L_{\mathrm{iso}}$ for \GRB\,
along with almost one hundred KW GRBs with known redshifts;
this sample is discussed in detail by Tsvetkova et al. (in preparation), hereafter T13.
One can see that \GRB\, lies at the upper edge of the $E_{\mathrm{iso}}$ distribution
and is nearly an order of magnitude more luminous than any burst at z$\sim$1.

In Figure~\ref{FigEisoLiso}, the KW detection horizon for a \GRB-like event is also indicated.
The limit of $z\sim$7.5 is estimated by applying the KW triggering algorithm
to simulated light curves at different redshifts.
The simulations, which account for both time dilation and spectral reddening,
show that at z$\geq5$, only the initial pulse remains detectable
in the KW band; at high redshifts the burst is seen as a short ($T_{90}\sim2-3$~s)
GRB with moderate $E_{\mathrm{peak}}\sim300$~keV and relatively long spectral lag of $\sim200$~ms.
Assuming a \SW/BAT sensitivity of $10^{-8}$~erg~cm$^{-2}$~s$^{-1}$ in the 15--150~keV band \citep{Barthelmy2005},
its detection threshold is reached at $z\sim$12.

The key rest-frame parameters of \GRB, the intrinsic peak energy, $E_{\mathrm{p,i}}\equiv E_{\mathrm{peak}}(1+z)=670\pm140$~keV, and
the rest-frame duration, $T_{90}/(1+z)=9.9\pm0.05$~s, are nearly at the center of the distributions for the KW sample and may be considered typical.
Consequently the huge energetics of \GRB\, cannot be easily explained
by an extremely hard spectrum or by long-lasting prompt emission.
Since more than half of the burst's energy is released in the hard initial pulse (Section~\ref{sec:tienerg}),
the intrinsic peak energy of the pulse's time-averaged spectrum ($2000\pm200$~keV) is also an important characteristic of the event.
It is in the top 10\% of the KW $E_{\mathrm{p,i}}$ distribution, but it cannot be the sole reason for the \GRB\, record energy.
Therefore, the most likely explanation for both the observed and isotropic-equivalent rest-frame energetics
is a huge photon flux, which suggests a highly-collimated emission.

\begin{figure*}[t!]
\centering
\includegraphics[width=0.9\textwidth]{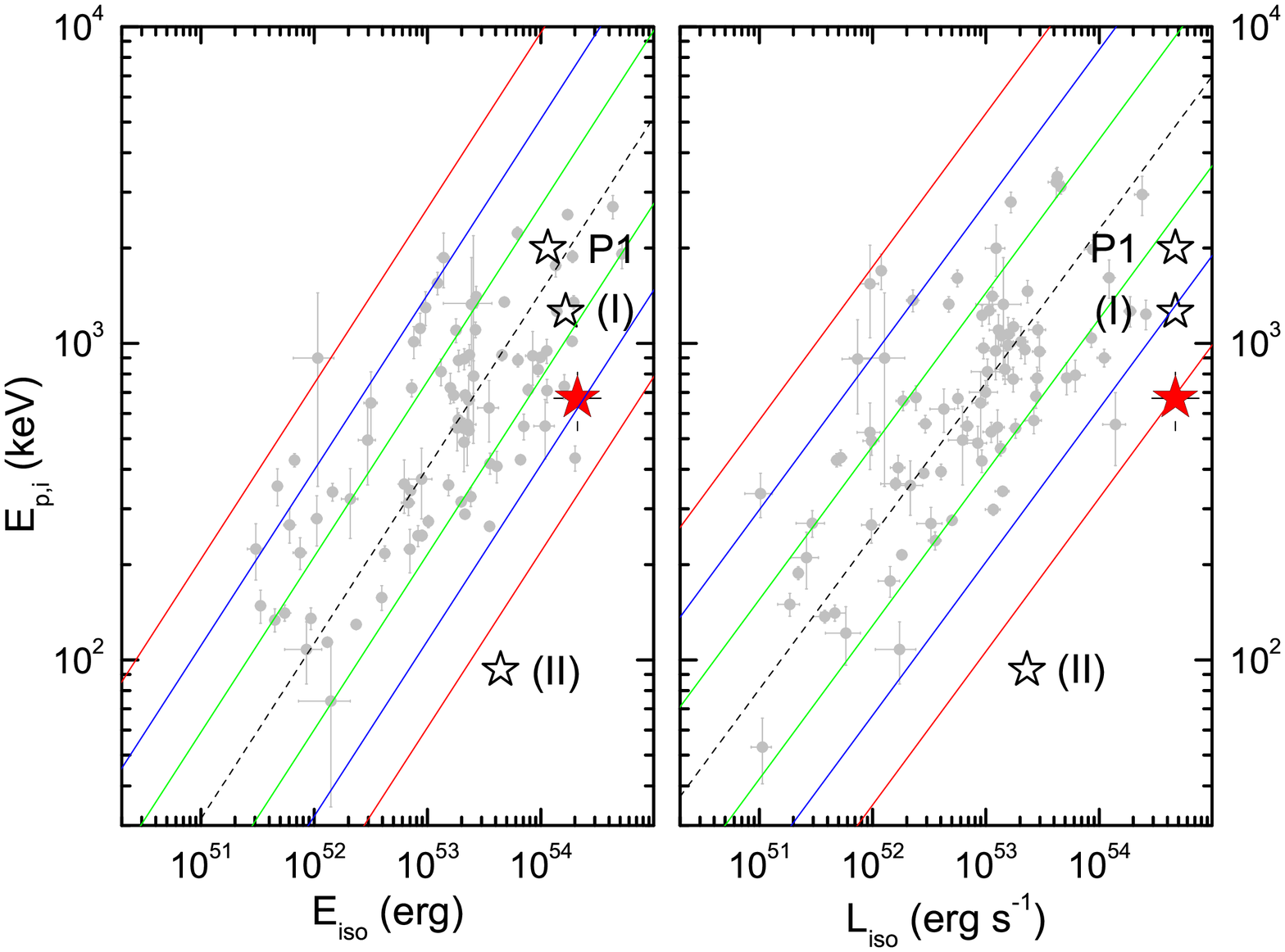}
\caption{
Rest-frame energetics in the $E_{\mathrm{iso}}-E_{\mathrm{p,i}}$ (left) and $L_{\mathrm{iso}}-E_{\mathrm{p,i}}$ (right)
planes. \GRB\, is shown with filled stars; the open symbols show Phase~I (I), Phase~II (II), and the initial pulse alone~(P1).
The KW GRBs with known redshifts are shown with gray symbols.
The recent updates for the Amati ($E_{\mathrm{p,i}}-E_{\mathrm{iso}}$) and Yonetoku ($E_{\mathrm{p,i}}-L_{\mathrm{iso}}$) relations from \cite{Ghirlanda2012}
are plotted with the dashed lines together with their 1$\sigma$, 2$\sigma$, and 3$\sigma$ scatters (solid lines).\\
}
\label{FigAmYn2}
\end{figure*}

We tested the \GRB\, rest-frame characteristics against $E_{\mathrm{p,i}}-E_{\mathrm{iso}}$ and $E_{\mathrm{p,i}}-L_{\mathrm{iso}}$
relations--known as the ``Amati'' and ``Yonetoku'' correlations (\citealt{Amati2002}; \citealt{Yonetoku2004}, respectively;
we used their recent updates from \citealt{Ghirlanda2012}).
One can see from Figure~\ref{FigAmYn2} that the burst is close to being a luminous outlier
with respect to both relations, suggesting that the spectral hardness of \GRB\,
is likely a minor ingredient of the burst's isotropic energy as compared to the population.
On the other hand, the spread in the spectral-energy correlations is known to diminish
when the jet effect is taken into account \citep{Ghirlanda2004,Ghirlanda2007};
this supports the hypothesis that the \GRB\, collimation is relatively tighter when compared to an average GRB.
It should be noted, finally, that the initial pulse taken alone follows the $E_{\mathrm{p,i}}-E_{\mathrm{iso}}$
relation to within 1$\sigma$ and the $E_{\mathrm{p,i}}-L_{\mathrm{iso}}$ relation to within $\sim1.5\sigma$.
This is particularly important since, as mentioned above, the initial pulse
is the only phase of the burst which could be seen from high redshifts and emphasizes
the role of observational bias in these studies.

\subsection{Bulk Lorentz Factor of the Ejecta}
\label{sec:lorentz}

Several methods have been proposed to estimate the bulk Lorentz factor of the GRB ejecta
($\Gamma_0$): the pair-opacity constraint from the ``compactness'' problem \citep{LS01},
the optical afterglow onset method (e.g., \citealt{SP99}),
and the very early external shock emission method, which considers
the quiescent periods between the prompt emission pulses \citep{ZP2010}.
Since no early observations are available for the \GRB\, afterglow
and the prompt emission pulses overlap considerably,
only the first method mentioned is applicable.

We calculated the lower limit on $\Gamma_0$ using inequalities from \cite{LS01}.
The first (Limit~A) is
\begin{equation}
 \Gamma_{\mathrm{0,A}} > \hat{\tau}^{1/(2\beta+2)}(E_{\max}/m_e c^2)^{(\beta-1)/(2\beta+2)}(1+z)^{{\beta-1}/{\beta+1}};
 \label{eq:LS01A}
\end{equation}
and the second (Limit~B) is
\begin{equation}
 \Gamma_{\mathrm{0,B}} > \hat{\tau}^{1/(\beta+3)}(1+z)^{{\beta-1}/{\beta+3}},
 \label{eq:LS01B}
\end{equation}
where $\beta$ is the photon power-law index at MeV energies (in the $N(E)\propto E^{-\beta}$ notation),
$E_{\max}$ is the highest energy at which photons were observed, and
\begin{equation}
 \hat{\tau} \simeq 4.3\times 10^{10} \frac{f_1}{\beta-1}\,  D_{L,28}^2 \, 0.511^{1-\beta}\, (\delta T/0.1)^{-1},
\end{equation}
where $\delta T$ is the minimum variability time scale of the prompt emission,
and $f_1$ is the photon flux at 1~MeV.

From the KW analysis, we assume the following parameters for \GRB:
$\delta T$=0.25~s, $\beta=2.39$, $E_{\mathrm{max}}>18$~MeV, and $f_1=160$~cm$^{-2}$~s$^{-1}$~MeV$^{-1}$,
all taken at the peak of the initial pulse (see Sections~\ref{sec:pulses} and \ref{sec:specres}).
This yields $\Gamma_{\mathrm{0,A}}>$240 and $\Gamma_{\mathrm{0,B}}>$360; by applying the tighter limit, we conclude $\Gamma_{0}>$360.

Typically, when photons are observed at high energies ($E_{\mathrm{max}}\sim$~GeV),
Limit A yields a stricter constraint than Limit B.
Unfortunately, \GRB\, was not observed by high-energy missions
and we proceeded from a very conservative \KW\, limit, $E_{\mathrm{max}}>18$~MeV.
However, the value of $\Gamma_{\mathrm{0,B}}$ allows us to set the lower limit on
the high-energy cutoff of the \GRB\, emission spectrum to $E_{\mathrm{max}}\geq100$~MeV
(assuming that the spectrum measured by \KW\, in the MeV band extends
unchanged to those energies).

\cite{Liang2010} discovered a tight correlation between $\Gamma_0$ and $E_{\mathrm{iso}}$.
This correlation was confirmed and refined on a broader GRB sample by \cite{Lu2012}
who obtained $\Gamma_0\simeq 91E^{0.29}_{\mathrm{iso,52}}$.
In the latter work, another tight correlation between $\Gamma_0$ and
$L_{\mathrm{iso}}$ was found in the form of $\Gamma_0\simeq 249L^{0.30}_{\mathrm{iso,52}}$.
Applied to \GRB\, these relations yield $\Gamma_0$ values of 430 and 1580, respectively.
Close values ($\Gamma_0(E_{\mathrm{iso}})\sim450$ and $\Gamma_0(L_{\mathrm{iso}})\sim1450$)
are evaluated based on slightly different slopes from \cite{Ghirlanda2012}.
It should be noted, however, that the GRB samples used by \cite{Lu2012}
and \cite{Ghirlanda2012} do not contain any burst with $\Gamma_0>1000$
or $L_{\mathrm{iso}}>10^{54}$~erg~s$^{-1}$, which is the case for this burst.
Thus, the $\Gamma_0$ value of $\sim450$ obtained from the $\Gamma_0-E_{\mathrm{iso}}$ relation,
which is tested by both authors up to $E_{\mathrm{iso}}\sim10^{54}$~erg,
may be considered more feasible than $\Gamma_0\sim1500$ derived from $L_{\mathrm{iso}}$.
The $\Gamma_0$ estimate of a few hundred is supported by another correlation
studied by \cite{Ghirlanda2012}, the $\Gamma_0-E_{\mathrm{p,i}}$ relation.
From $E_{\mathrm{p,i}}=670$~keV, this relation yields $\Gamma_0=150$ and,
from $E_{\mathrm{p,i}}=2000$~keV for the initial pulse, $\Gamma_0=400$.

\subsection{Scenarios for Afterglow and Extended $\gamma$-ray Emission}
\label{sec:closure}

\begin{figure}
\centering
\includegraphics[width=0.5\textwidth]{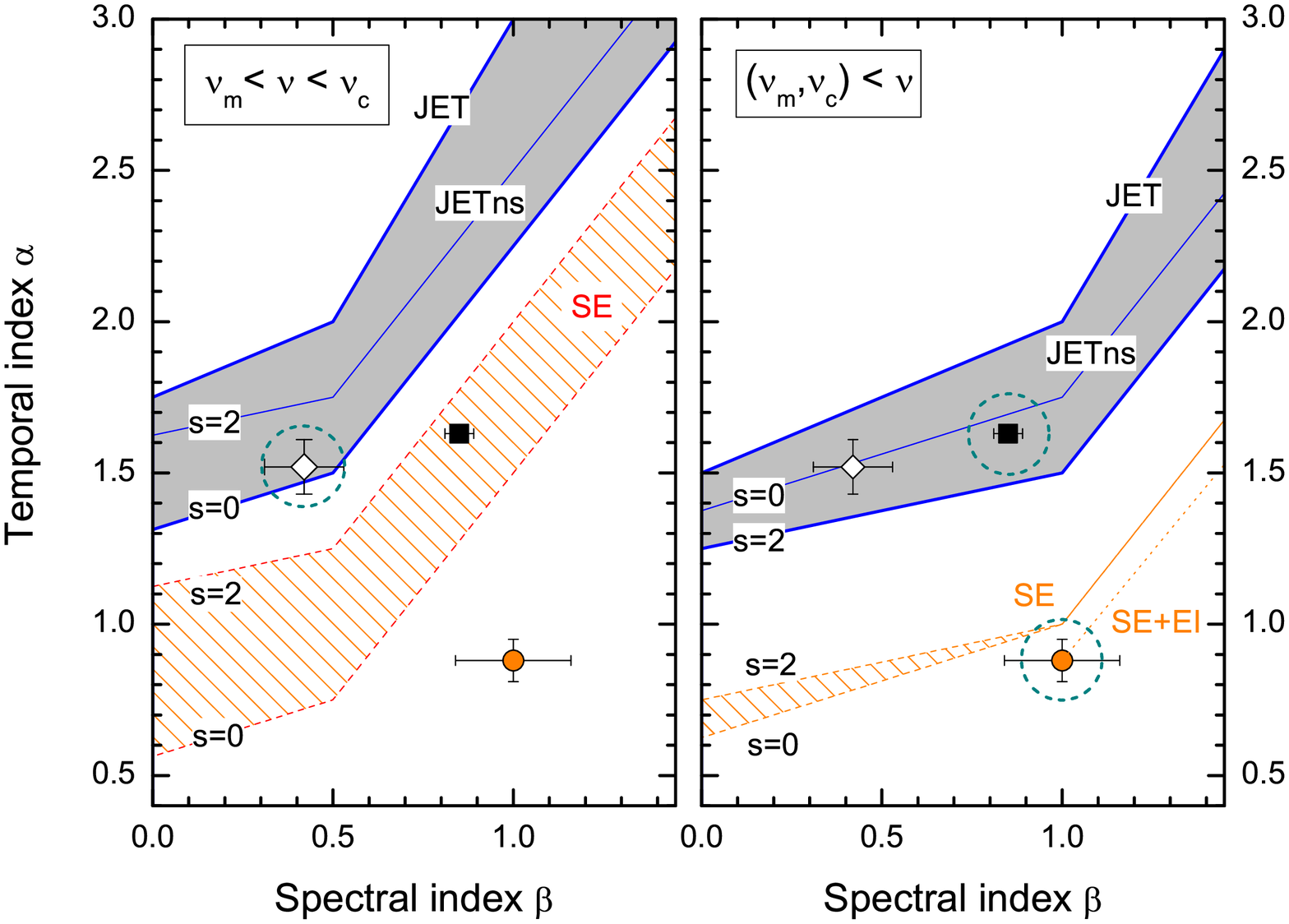}
\caption{Closure relations (CRs) of the synchrotron forward-shock model.
Left and right panels are for the $\nu_m<\nu<\nu_c$ and $\nu_m,\nu_c<\nu$ regimes, respectively.
The solid lines and the solid shaded regions indicate CRs for the post-break segment IV of the canonical light curve.
The upper and lower boundaries of the solid regions are defined for
a jet with (JET) and without (JETns) sideways expansion taken into account, respectively.
The dashed lines and hatched regions indicate CRs for the pre-break, spherical geometry (SE).
SE accompanied by late-time energy injection (EI) with $q=0.88$ is shown with the dotted line.
Boundary cases for the CBM density profile $n(r)\propto r^{-s}$ are indicated, $s=0$ (ISM) and $s=2$ (wind).
The small circles stand for the extended $\gamma$-ray emission and the squares for the X-ray afterglow.
The UVOT temporal index is shown in combination with the spectral index obtained from the
broadband SED (open diamonds).
The large dashed circles indicate the ejecta geometry and the cooling regime we consider
to be likely for the observations (see Section~\ref{sec:closure} for the discussion).
}
\label{FigCR}
\end{figure}

\begin{deluxetable*}{lccccclc}
\tablewidth{0pt}
\tablecaption{Closure Relations
\label{TableCR}}
\tabletypesize{\footnotesize}
\tablehead{
\colhead{Geometry} & \colhead{Medium} & \colhead{Spectral Regime} & \colhead{$p$} & \colhead{$\alpha_{\mathrm{cr}}$}& \colhead{$f_{\mathrm{cr}}$} &  \colhead{CR} &  \colhead{Reference}
}
\startdata
\multicolumn{8}{c}{Extended $\gamma$-ray emission ($\alpha_{\mathrm{obs}}=0.88\pm0.05$, $\beta_{\mathrm{obs}}=1.00\pm0.16$) } \\[0.1cm]
\cline{1-8}
$SE$ & $ISM$          & $\nu_{\gamma}<\nu_c$   &   $3.0\pm0.3$  & 1.5 & 3.3$\sigma$    & $\alpha=3\beta/2$ &1a    \\     
$SE$ & $Wind$         & $\nu_{\gamma}<\nu_c$   &   $3.0\pm0.3$  & 2.0 & 6.0$\sigma$    & $\alpha=(3\beta+1)/2$ &5a    \\     
$SE$ & $ISM, Wind$    & $\nu_c<\nu_{\gamma}$   &   $2.0\pm0.3$  & 1.0 & 0.6$\sigma$    & $\alpha=(3\beta-1)/2$ &2a,6a \\     
\cline{1-8}
\multicolumn{7}{c}{XRT ($\alpha_{\mathrm{obs}}=1.63\pm0.02$, $\beta_{\mathrm{obs}}=0.85\pm0.04$) } \\
\cline{1-8}
$SE$ & $ISM$          & $\nu_{\mathrm{X}}<\nu_c$          &   $2.7\pm0.1$  & 1.28 & 5.6$\sigma$    & $\alpha=3\beta/2$ &1a   \\     
$SE$ & $Wind$         & $\nu_{\mathrm{X}}<\nu_c$          &   $2.7\pm0.1$  & 1.83 & 2.3$\sigma$    & $\alpha=(3\beta+1)/2$ &5a   \\     
$JET$ &$ISM, Wind$ & $\nu_{\mathrm{X}}<\nu_c$          &   $2.7\pm0.1$  & 2.70 &  13$\sigma$       & $\alpha=2\beta+1$ &9a   \\     
$JETns$ & $ISM$     & $\nu_{\mathrm{X}}<\nu_c$          &   $2.7\pm0.1$  & 2.03 & 6.2$\sigma$      & $\alpha=(6\beta+3)/4$ &11a  \\     
$JETns$ & $Wind$    & $\nu_{\mathrm{X}}<\nu_c$          &   $2.7\pm0.1$  & 2.28 & 10 $\sigma$      & $\alpha=(3\beta+2)/2$ &13a  \\     
\\
$SE$ & $ISM$          & $\nu_c<\nu_{\mathrm{X}}$          &   $1.7\pm0.1$  & 0.94 & 27$\sigma$     & $\alpha=(3\beta+5)/8$ &2b   \\     
$SE$ & $Wind$         & $\nu_c<\nu_{\mathrm{X}}$          &   $1.7\pm0.1$  & 0.96 & 30$\sigma$     & $\alpha=(\beta+3)/4$ &6b   \\     
$JET$ &$ISM, Wind$ & $\nu_c<\nu_{\mathrm{X}}$          &   $1.7\pm0.1$  & 1.93 & 10$\sigma$        & $\alpha=(\beta+3)/2$ &10b  \\     
$JETns$ & $ISM$     & $\nu_c<\nu_{\mathrm{X}}$          &   $1.7\pm0.1$  & 1.69 & 2.5$\sigma$      & $\alpha=(3\beta+11)/8$ &12b  \\     
$JETns$ & $Wind$    & $\nu_c<\nu_{\mathrm{X}}$          &   $1.7\pm0.1$  & 1.46 & 7.5$\sigma$      & $\alpha=(\beta+5)/4$ &14b  \\     
\cline{1-8}
\multicolumn{7}{c}{UVOT ($\alpha_{\mathrm{obs}}=1.52\pm0.09$, $\beta_{\mathrm{obs}}=0.42\pm0.11$) } \\[0cm]
\cline{1-8}
$SE$ & $ISM$          & $\nu_{\mathrm{opt}}<\nu_c<\nu_{\mathrm{X}}$      &   $1.8\pm0.2$  & 0.72 & 8.1$\sigma$    & $\alpha=3(2\beta+3)/16$ &1b   \\     
$SE$ & $Wind$         & $\nu_{\mathrm{opt}}<\nu_c<\nu_{\mathrm{X}}$      &   $1.8\pm0.2$  & 1.23 & 3.1$\sigma$    & $\alpha=(2\beta+9)/8$ &5b   \\     
$JET$ &$ISM, Wind$ & $\nu_{\mathrm{opt}}<\nu_c<\nu_{\mathrm{X}}$      &   $1.8\pm0.2$  & 1.96    & 9.1$\sigma$       & $\alpha=(2\beta+7)/4$ &9b   \\     
$JETns$ & $ISM$     & $\nu_{\mathrm{opt}}<\nu_c<\nu_{\mathrm{X}}$      &   $1.8\pm0.2$  & 1.47   & 0.5$\sigma$      & $\alpha=(6\beta+21)/16$ &11b  \\     
$JETns$ & $Wind$    & $\nu_{\mathrm{opt}}<\nu_c<\nu_{\mathrm{X}}$      &   $1.8\pm0.2$  & 1.73   & 2.2$\sigma$      & $\alpha=(2\beta+13)/8$ &13b  \\     
\enddata
%
    \tablecomments{ \footnotesize{
    Column 1 is the outflow geometry and dynamics: $SE$ (spherical expansion),
    $JET$ (spreading jet), $JETns$ (non-spreading jet).
    Column 2 is the circum-burst density profile $n(r)\propto r^{-s}$: $ISM$ ($s=0$), $Wind$ ($s=2$).
    Column 3 is the position of the observation band $\nu$ relative to the synchrotron cooling frequency $\nu_c$;
    the slow-cooling regime, $\nu_m<min(\nu,\nu_c)$, is assumed.
    Column 4 is the electron spectral distribution index, given the spectral regime and $\beta_{\mathrm{obs}}$.
    Column 5 is the CR-predicted temporal index $\alpha_{\mathrm{cr}}$.
    Column 6 is the deviation of the observed temporal index $\alpha_{\mathrm{obs}}$ from $\alpha_{\mathrm{cr}}$.
    Column 7 is the closure relation and column (8) is its index in Table~1 of \cite{Racusin2009},
    where the references can be found.
    }
   }

\end{deluxetable*}

In Section~\ref{sec:extended}, it was shown that at $\sim$30~s after the KW trigger,
the prompt phase of \GRB\, rapidly developed into a steadily decaying
``tail'' of the extended $\gamma$-ray emission. The decay follows a power law with
the temporal index $\alpha_{\gamma}\approx0.88$ out to $\sim~T_0+700$~s.
At the same time, the emission spectrum is well described by
a power law with spectral index $\beta_{\gamma}\approx1$.
The combination of the spectral and temporal behavior is in fairly good agreement with
that expected at the ``plateau'' (II) and ``normal'' (III)
phases of the canonical X-ray afterglow \citep{Nousek2006,Zhang2006}.
This supports a scenario in which the extended $\gamma$-ray emission of \GRB\,
is generated by the synchrotron forward-shock mechanism with a possible late-time energy injection.

\SW\, observations of the \GRB\, afterglow started $\sim107$~ks after the burst
and showed no clear signature of a possible jet break in the X-ray and UV/optical
light curves up to the time the observations ceased $\sim$48~days later.
The XRT temporal slope, $\alpha_{\mathrm{X}}\approx1.63$, cannot be unambiguously attributed to the
``normal'' (III) or ``jet'' (IV) segment of the canonical X-ray afterglow;
$\alpha_{\mathrm{X}}$ is at the steep end of the segment III indices and at
the shallow end of the segment IV indices for a sample of XRT afterglows with
``prominent'' jet breaks \citep{Racusin2009}.
The slope of 1.5, which is widely suggested as critical to distinguish
the pre-break and post-break segments of both X-ray and optical afterglows
(see, e.g., \cite{Panaitescu2007}, \cite{Liang2008}), is also too close to
$\alpha_{X}$ and $\alpha_{\mathrm{opt}}\approx1.52$ to draw any firm conclusion.
Thus the question of the jet break location with respect to the time span of the \SW\,
observations cannot be easily answered when discussing the afterglow temporal slopes alone.

The standard fireball model for GRB afterglows suggests that their temporal $\alpha$,
and spectral $\beta$, indices follow closure relations (CRs),
which are linear relationships between $\alpha$ and $\beta$,
both being linear functions of a power-law index of the underlying electron spectrum $p$.
We tested the \GRB\, observations against an extensive set of CRs
collected in Table~1 of \cite{Racusin2009}, which, in turn, are taken from
\cite{ZM2004}, \cite{DC2001}, \cite{Zhang2006}, \cite{Panaitescu2005}, and \cite{Panaitescu2006}.
Following \cite{Zhang2006}, we expect the typical synchrotron frequency, $\nu_m$,
to lie below the observational bands and, for the XRT and UVOT observations,
which started $\sim107$~ks after the burst,
we assume the slow-cooling regime ($\nu_m<\nu_c$),
where $\nu_c$ is the synchrotron-cooling frequency.

In Table~\ref{TableCR}, the values of $p(\beta_{\mathrm{obs}})$ and the deviations
$f_{\mathrm{cr}}\equiv\alpha_{\mathrm{cr}}(\beta_{\mathrm{obs}})-\alpha_{\mathrm{obs}}$ are listed, where $\alpha_{\mathrm{cr}}$ is
the temporal slope predicted by the closure relation and the indices $\alpha_{\mathrm{obs}}$, $\beta_{\mathrm{obs}}$,
are obtained from our observations.
Figure~\ref{FigCR} shows the $\alpha(\beta)$ diagram for the closure relations and the data.

\emph{X-ray afterglow.} In the $\nu_c<\nu_{\mathrm{X}}$ regime, the electron index $p=2\beta_{\mathrm{X}}=1.7\pm0.14$.
From $(\alpha_{\mathrm{X}},\beta_{\mathrm{X}})$ and CRs for this spectral regime (Figure~\ref{FigCR}, right)
a pre-break geometry of ejecta which mimics an isotropic, spherical expansion (SE),
is rejected at $>20\sigma$.
The best-fitting CR (2.5$\sigma)$ is for the non-spreading jet (JETns) and ISM-like circum-burst medium (CBM).
Another option, the $\nu_{\mathrm{X}}<\nu_c$ spectral regime, which is considered unlikely for late-time X-ray afterglows
(see, e.g., \cite{Zhang2006,Nysewander2009}), yields a steep electron spectrum, $p=2\beta_{\mathrm{X}}+1=2.7\pm0.14$.
In this case, all the jet closure relations are $>6\sigma$ from the X-ray observations;
the best fitting CR suggests the pre-break (SE) geometry and a stellar-wind-like CBM ($2.3\sigma$).

\emph{UV/optical afterglow.} The UVOT temporal slope, $\alpha_{\mathrm{opt}}=1.52\pm0.09$,
is shallower but consistent with $\alpha_{X}$.
For the spectral slope in the optical band, we accept $\beta_{\mathrm{opt}}=0.42\pm0.11$
(1$\sigma)$ obtained from the broadband XRT+UVOT+GROND SED in Section~\ref{sec:sed}.
The $\nu_c<\nu_{\mathrm{opt}}$ regime implies an ultra-hard electron index
of $p=2\beta_{\mathrm{opt}}=0.84\pm0.22$, which is at odds with both values of $p$ suggested from the X-ray spectrum.
Alternatively, in the $\nu_{\mathrm{opt}}<\nu_c$ regime $p=2\beta_{\mathrm{opt}}+1=1.84\pm0.22$,
which is in good agreement with the electron index found for $\nu_c<\nu_{\mathrm{X}}$.
In this case, a cooling break of $\Delta\beta=1/2$ is expected between the UVOT and XRT bands,
which is consistent with $\beta_{X}-\beta_{\mathrm{opt}}=0.43\pm0.12$.
From $(\alpha_{\mathrm{opt}},\beta_{\mathrm{opt}})$ all pre-break CRs are $>3\sigma$ from
the data and the JETns+ISM is again the best fitting CR ($0.5\sigma$).

\emph{Extended $\gamma$-ray emission.} Given $\beta_{\gamma}=1.00\pm0.16$
and $\alpha_{\gamma}=0.88\pm0.05$,
the fast-cooling regime $\nu_c<\nu_{\gamma}<\nu_m$, which requires $\beta=1/2$
for the ISM and $\alpha=0$ for the wind-like CBM, is ruled out at $>3\sigma$.
The same is true for the slow-cooling regime and $\nu_{\gamma}<\nu_c$
(see Figure~\ref{FigCR}, left panel).
In the $max(\nu_m,\nu_c)<\nu_{\gamma}$ regime, the electron index is
$p=2\beta_{\gamma}=2.0\pm0.3$, which is consistent with $p\sim1.7$,
implied for $\nu_c<\nu_{\mathrm{X}}$. The SE closure relation fits the observations
within $1\sigma$ (same Figure, right panel), with $\alpha_{\gamma}$ being
slightly shallower than the value of unity predicted by the CR.
This shallow slope could be explained by a continuous energy injection (EI),
caused, e.g., by prolonged activity of a central engine \citep{ZM2001,Zhang2006}.
The EI can be characterized by a parameter $q<1$ such that the continuous
luminosity injection is $L(t)\propto t^{-q}$; $q=1$ means no additional energy injection.
The CR for EI in the $max(\nu_m,\nu_c)<\nu$ regime is $\alpha=(q-2)/2 +\beta(2+q)/2$ \citep{Zhang2006}.
This yields $q=0.88\pm0.15$ and suggests a mild late-time energy injection.

Thus, the CR analysis results in a consistent scenario
for the extended $\gamma$-ray emission, X-ray, and UV/optical afterglows.
The scenario is characterized by a hard electron spectrum,
with $p$ in the 1.7--2.0 range, and implies the $\nu_{\mathrm{opt}}<\nu_c<(\nu_{X},\nu_{\gamma})$
spectral regime for the observations.
In this scenario, the X-ray and UV/optical afterglows are observed
after the jet break and, assuming a forward-shock origin,
the extended $\gamma$-ray emission is generated at the pre-break phase of the fireball expansion.
These choices are indicated by dashed circles in Figure~\ref{FigCR}.
The temporal slope change $\alpha_{X}-\alpha_{\gamma}=0.75\pm0.05$ is
in reasonably good agreement ($<2\sigma$) with the jet break,
$\Delta \alpha=2/3$, expected for the combination of $p\leq2$,
non-spreading jet, and ISM-like environment;
when accounting for the late-time energy injection, the agreement is even better.
The shallow post-break temporal slope can be explained, in this model,
by a ``flat'' electron spectrum and by a limited lateral spreading of the jet.

We note, that these conclusions are drawn under the assumption of
non-transitional regimes for the cooling and the geometry;
this approach is justified by the results of our temporal and spectral analysis
of the extended emission and the afterglow.

\subsection{Extrapolated X-ray Light Curve and Prompt Emission--Afterglow Correlations}
\label{sec:extrap}
\begin{figure}
\centering
\includegraphics[width=0.5\textwidth]{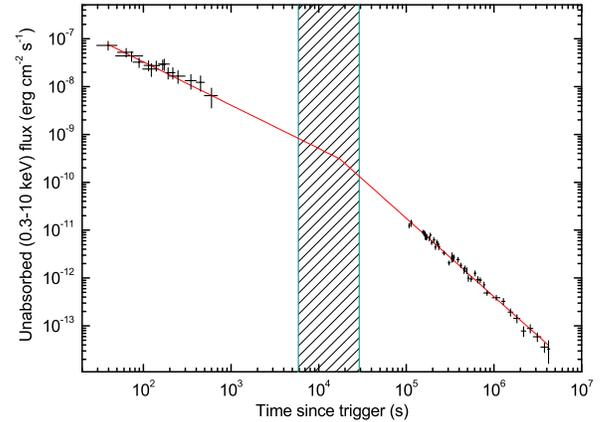}
\caption{ Hypothetical light curve of the \GRB\, X-ray afterglow
constructed from the \KW, SPI-ACS, and XRT observations.
Data points prior to $T_0+1000$~s are extrapolated from the KW and SPI-ACS
data (Section~\ref{sec:extrap}). The best BPL fit is shown by the solid line
and the 1$\sigma$ break region $(1.74\pm1.15)\times10^4$~s is indicated by the hatched area.
}
\label{FigKWACSXRTfit}
\end{figure}

Assuming that the power-law energy spectrum measured by \KW\, at $>T_0+30$~s
extends unchanged to lower energies, the extrapolated X-ray flux light curve
can be calculated from the extended $\gamma$-ray emission.
The correctness of this approach has been tested, e.g., with the KW data on the ``naked-eye'' GRB~080319B
and the recent nearby GRB~130427A, for which statistically significant tails of
an extended emission in the 20--1200~keV band have been observed,
both having power-law spectra with $\Gamma\sim2$.
For both GRBs, we found the extrapolated unabsorbed 0.3--10~keV flux to be consistent,
within 2-3$\sigma$ (or $\leq$20\%),
with simultaneous \SW-XRT observations extracted from the XRT light curve repository \citep{Evans2007,Evans2009}.

Using the KW counts-to-flux conversion factor and the spectral index from the $T_0+50\,-\,T_0+200$~s interval,
we found the extrapolated unabsorbed 0.3--10~keV flux to be in the $(10^{-8}\,-\,10^{-7})$~erg~cm$^{-2}$~s$^{-1}$ range,
which has been previously reported at early phases of bright X-ray afterglows
(e.g., GRB~061007, \cite{Schady2007};  GRB~080319B, \cite{Racusin2008}; GRB~130427A, \cite{GCN14502,Perley2013}).
In Figure~\ref{FigKWACSXRTfit}, a hypothetical light curve, which combines the extrapolated
late-time $\gamma$-ray observations of \GRB\, and the XRT afterglow, is shown.
Assuming that the accuracy of the extrapolation procedure is comparable
to that found for GRB~080319B and GRB~130427A, we added 20\% systematics to the statistical-only errors
for the extrapolated points.
A transition from the initial, flat slope to the final, steeper one occurs in
the $T_0+10^3\,-\,T_0+10^5$~s region, which is not covered by the observations.
A simple estimate from the BPL fit to this combined light curve yields,
at $\chi^2_r=0.97$, the 1$\sigma$ break region of $t_b=(1.74\pm1.15)\times10^4$~s
with pre- and post-break slopes of $\alpha_1=0.90\pm0.09$ and $\alpha_2=1.64\pm0.02$,
respectively.

In the post-break scenario for the X-ray afterglow, $t_b\sim1.7\times10^4$~s (or $\sim0.2$~d)
may be considered as a lower limit for $t_{\mathrm{jet}}$,
since the extended $\gamma$-ray emission cannot be attributed unambiguously to
a ``plateau'' (II) or ``normal'' (III) phase.
In the former case, introducing an additional, ``injection'' break to the light curve
(the break from segment~II to segment~III) will shift the jet break
towards the start of the XRT observations at $\sim107$~ks (1.24~d) after the burst.
The 0.2--1.24~d range for $t_{\mathrm{jet}}$ is consistent with the value suggested
from the multi-parameter $E_{\mathrm{iso}}-E_{\mathrm{p,i}}-t_{\mathrm{jet}}$ correlation
\citep{LZ2005,Ghirlanda2007}, which predicts, for \GRB, the jet break
at 0.3$\pm$0.13~d using the slopes provided in the latter work.

The isotropic-equivalent $K$-corrected X-ray luminosity in the rest-frame energy band $E_1-E_2$ can be calculated from the unabsorbed 0.3--10~keV flux $F_{\mathrm{X}}\propto E^{-\beta_{\mathrm{X}}}$:
\begin{equation}
L_{\mathrm{X}}(t)=4\pi D_L^2 \frac{F_{\mathrm{X}}[t(1+z)]}{(1+z)^{1-\beta_{\mathrm{X}}}} \frac{(E_2^{1-\beta_{\mathrm{X}}}-E_1^{1-\beta_{\mathrm{X}}})}{(10^{1-\beta_{\mathrm{X}}}-0.3^{1-\beta_{\mathrm{X}}})},
\end{equation}
where $E_1$ and $E_2$ are measured in keV.
The \GRB\, afterglow luminosities at 5 minutes, 1 hour, 11 hours, and one day after the trigger in the rest-frame
were estimated from the combined extrapolated light curve and are listed in Table~\ref{TableLx}.
We tested the 2--10~keV $L_{\mathrm{X}}$ and the rest-frame prompt emission parameters against
$L_{\mathrm{X}}-E_{\mathrm{iso}}$, $L_{\mathrm{X}}-L_{\mathrm{iso}}$, and $L_{\mathrm{X}}-E_{\mathrm{p,i}}$ correlations reported by \cite{D'Avanzo2012}
from a sample of 56 bright \SW\, GRBs with well-measured X-ray afterglows.
\GRB\, was found to fit these relations within 3$\sigma$ scatter at all four time marks.
The best fit ($\leq1\sigma$) is for the $L_{\mathrm{X}}-E_{\mathrm{p,i}}$ relation;
the data are consistently $\sim2.5-3\sigma$ from the $L_{\mathrm{X}}-L_{\mathrm{iso}}$ relation,
in general agreement with other correlations for which the extreme
$L_{\mathrm{iso}}$ of this GRB is considered.

\begin{deluxetable}{lcccc}
\tablewidth{0pt}
\tablecaption{X-ray Afterglow Luminosity
\label{TableLx}}
\tablehead{
\colhead{Band}  & \colhead{$L_{X,5}$ } & \colhead{$L_{X,1}$} & \colhead{$L_{X,11}$} & \colhead{$L_{X,24}$}
\\
\colhead{(keV)} & \colhead{(erg~s$^{-1}$)} & \colhead{(erg~s$^{-1}$)} & \colhead{(erg~s$^{-1}$)}  & \colhead{(erg~s$^{-1}$)}
}
\startdata
0.3--10         & $1.5\times 10^{49}$      & $1.7\times 10^{48}$      & $3.5\times 10^{46}$       & $1.0\times 10^{46}$  \\
2--10           & $6.9\times 10^{48}$      & $7.8\times 10^{47}$      & $1.8\times 10^{46}$       & $5.2\times 10^{45}$  \\
\enddata
    \tablecomments{ \footnotesize{
    Columns 2 to 4 are the luminosities at 5 minutes, 1 hour, 11 hours, and one day after the trigger in the rest-frame, respectively.
    }
   }
\end{deluxetable}

\subsection{Jet Opening Angle and Collimation-corrected Energy}
\label{sec:jet}

In the case of an ISM-like CBM
with constant number density $n$, the jet opening angle is given by \cite{SPH1999}:
\begin{equation}
\theta_{\mathrm{jet}}=\frac{1}{6}\left(\frac{t_{\mathrm{jet}}}{1+z}\right)^{3/8}\left(\frac{n\eta_{\gamma}}{E_{\mathrm{iso,52}}}\right)^{1/8},
\end{equation}
where $\eta_{\gamma}$ is the radiative efficiency and $t_{\mathrm{jet}}$ is measured in days.

\begin{deluxetable}{lcccc}
\tablewidth{0pt}
\tablecaption{Collimation-corrected energy
\label{TableJet}}
\tablehead{
\colhead{$t_{\mathrm{jet}}$} & \colhead{$\theta_{\mathrm{jet}}$ } & \colhead{Collimation} & \colhead{$E_{\gamma}$} & \colhead{$L_{\gamma}$}
\\
\colhead{(days)} & \colhead{(deg)} & \colhead{factor}             & \colhead{(erg)}  & \colhead{(erg~s$^{-1}$)}
}
\startdata
0.20  &     1.7    & $4.4\times10^{-4}$     & $9.2\times10^{50}$      & $2.1\times10^{51}$       \\
1.24  &    3.4     & $1.7\times10^{-3}$     & $3.6\times10^{51}$      & $8.1\times10^{51}$      \\
$>$48 &    $>$13 & $>2.7\times10^{-2}$    & $>5.6\times10^{52}$     & $>1.2\times10^{53}$       \\
\enddata
\tablecomments{ \footnotesize{
The estimates are given for an ISM-like CBM with constant number density $n=1$~cm$^{-3}$; the radiative efficiency $\eta_{\gamma}=0.2$ has been assumed.
}}
\end{deluxetable}

For calculations, we adopted canonical values $\eta_{\gamma}=0.2$ and $n=1$~cm$^{-3}$ \citep{Frail2001}.
In Table~\ref{TableJet}, values of $\theta_{\mathrm{jet}}$, the collimation factor $(1-\cos(\theta_{\mathrm{jet}}))$,
the collimation-corrected energy release in $\gamma$-rays $E_{\gamma}=E_{\mathrm{iso}}(1-\cos(\theta_{\mathrm{jet}}))$,
and the collimation-corrected peak luminosity $L_{\gamma}=L_{\mathrm{iso}}(1-\cos(\theta_{\mathrm{jet}}))$
are listed for the key points of the \GRB\, observations timeline:
$t_{\mathrm{jet}}=1.2$~d (start of the \SW\, observations),
$t_{\mathrm{jet}}=48$~d (end of the \SW\, observations),
and $t_{\mathrm{jet}}=0.2$~d (a lower limit for the jet break suggested in the previous section).

\begin{figure}[t!]
\centering
\includegraphics[width=0.5\textwidth]{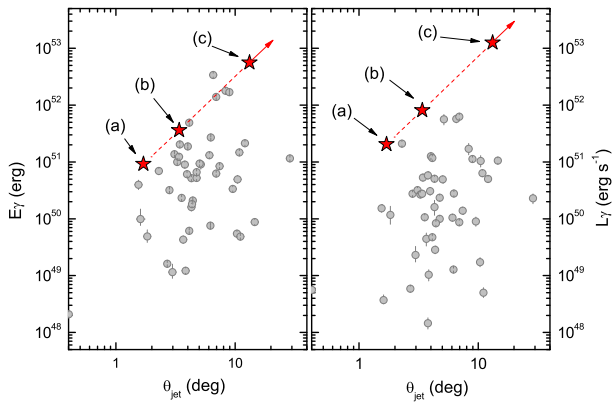}
\caption{
Collimation-corrected energy of the \GRB\, prompt emission.
The collimated energy release $E_{\gamma}$ (left) and the collimated peak luminosity $L_{\gamma}$
(right) are plotted \emph{vs.} the collimation angle.
Filled stars (a), (b), and (c) show the \GRB\, energies for an ISM
and $t_{\mathrm{jet}}$ at 0.2~d, 1.24~d (start of \SW\, observations), and 48~d
(end of \SW\, observations), respectively (see Table~\ref{TableJet}).
gray circles indicate KW GRBs with known collimation factors (Tsvetkova et al., in preparation).\\[0.2cm]
}
\label{FigEgLg}
\end{figure}

Figure~\ref{FigEgLg} shows our estimates of the \GRB\,
collimation-corrected energy.
For the jet-break time favored from our previous analysis,
between 0.2 and 1.24~days (points (a) and (b) in the figure),
the implied collimation angle is small, 1.7$\arcdeg$-3.4$\arcdeg$,
and the inferred values of $E_{\gamma}$ and $L_{\gamma}$ are
in the ranges determined by T13 for the KW GRBs with known collimation factors;
also, the $E_{\gamma}$ range is effectively the $_{-1.5}^{+2.5}\sigma$
scatter of the $E_{\mathrm{p,i}}-E_{\gamma}$ relation \citep{Ghirlanda2007}
calculated at $E_{\mathrm{p,i}}$=670~keV.
It should be noted, however, that among such tightly collimated bursts,
the fraction of which we estimate to be $<25$\% of KW GRBs with known collimation factors,
both $E_{\gamma}$ and $L_{\gamma}$ for \GRB\, are at the upper edges of their distributions;
this implies that the burst's intrinsic brightness is still remarkable.
In the case of a ``hidden'' break in the afterglow light curve, between points (b) and (c) in Figure~\ref{FigEgLg},
the inferred radiated energy still does not move far beyond the observed
range $E_{\gamma}\leq3.4\times10^{52}$~erg~s$^{-1}$.
However, the collimated peak luminosity shifts toward
an unprecedented $L_{\gamma}\sim10^{53}$~erg~s$^{-1}$,
an order of magnitude higher than the current record.

For a stellar-wind-like environment with $n(r)=5\times10^{11}A_*r^{-2}$,
the jet opening angle depends on $t_{\mathrm{jet}}$ through \citep{CL2000}:
\begin{equation}
\theta_{\mathrm{jet}}=0.202\left(\frac{t_{\mathrm{jet}}}{1+z}\right)^{1/4}\left(\frac{A_*\eta_{\gamma}}{E_{\mathrm{iso,52}}}\right)^{1/4},
\end{equation}
where $A_*=(\dot{M}_W/4\pi V_W)/5\times10^{11}$~g~cm$^{-1}$ is the wind parameter,
$\dot{M}_W$ is the mass-loss rate due to the wind, and $V_W$ is the wind velocity;
$A_*\sim1$ is typical for a Wolf-Rayet star.
Although the wind-like CBM is not preferred from our analysis,
the $\theta_{\mathrm{jet}}$ estimate for $A_*=1$ also results in a tight collimation
angle $\sim4.5\arcdeg$ and reasonable estimates for $E_{\gamma}\sim6.5\times10^{51}$~erg
and $L_{\gamma}\sim1.5\times10^{52}$~erg~s$^{-1}$ even at $t_{\mathrm{jet}}=$48~d.

\subsection{\GRB\, and the Most Luminous GRBs}
\label{sec:grbs}

\begin{deluxetable*}{lccccccc}
\tablewidth{0pt}
\tablecaption{The Most Luminous GRBs
\label{TableGRBs}}
\tablehead{
\colhead{Parameter} & \colhead{GRB~110918A\tablenotemark{a}} & \colhead{GRB~130505A} & \colhead{GRB~050603} & \colhead{GRB~080916C} & \colhead{GRB~080607} & \colhead{GRB~080721} & \colhead{GRB~000131}
}
\startdata
z (reference)                               & 0.984 (1)   & 2.27 (2)              & 2.821 (3)            & 4.35 (4)              & 3.036 (5)            & 2.591 (6)            & 4.511 (7)\\
$L_{\mathrm{iso}}$ ($10^{54}$~erg~s$^{-1}$) & 4.7         & 2.7                   & 2.6                  & 2.4                   & 1.9                  & 1.2                  & 1.1\\
$E_{\mathrm{iso}}$ ($10^{54}$~erg)& 2.1 [1.2]             & 3.8                   & 0.85                 & 4.3                   & 2.0                  & 1.3                  & 1.6\\
$E_{\mathrm{p,i}}$ (keV)          & 670 [2000]            & 2060                  & 1100                 & 2700                  & 1350                 & 1760                 & 700\\
$T_{90}/(1+z)$ (s)                & 9.9 [1.3]             & 6.6                   & 2.9                  & 11.5                  & 7.1                  & 5.4                  & 17.5
\enddata
\tablenotetext{a}{Values in square parentheses are for the \GRB\, initial pulse, the only
phase of the burst which could be seen from high redshifts.}
\tablerefs{(1) \citealt{Elliott2013}; (2) \citealt{GCN14567}; (3) \citealt{GCN3520}; (4) \citealt{Greiner2009}; (5) \citealt{Prochaska2009}; (6) \citealt{Starling2009}; (7) \citealt{Andersen2000}.}
\end{deluxetable*}

To date, seven $\gamma$-ray bursts with $L_{\mathrm{iso}}>10^{54}$~erg~s$^{-1}$ have been observed;
Table~\ref{TableGRBs} lists estimates of their rest-frame prompt emission characteristics
derived from the \KW\, observations.
While the scope of this paper does not involve an analysis of this sample,
one can see that the \GRB\, parameters do not differ much from those of other ultraluminous events:
almost all of them are hard-spectrum GRBs with a moderately short rest-frame duration.

Notably, none of these bursts (with the exception of GRB~000131, \citealt{Andersen2000})
demonstrate a ``canonical'' steep late-time decay of an afterglow light curve
\citep{Grupe2006,Greiner2009,Prochaska2009,Starling2009}.
When the late-time temporal and spectral power-law indices
of X-ray afterglows are considered, the second and third most luminous GRBs
(GRB~130505A\footnote[1]{Results of the XRT automated analysis \citep{Evans2007,Evans2009}
can be found at \url{http://www.swift.ac.uk/xrt\_products/00555163}}
and GRB~050603 \citealt{Grupe2006}) resemble \GRB\, most closely,
suggesting a similar interpretation. Particularly, a flat electron energy
spectrum with $p\sim1.4$ is implied for GRB~050603 by \cite{Grupe2006},
as well as a jet scenario for its afterglow observed by \SW\ from 11~hr after
the trigger onward. For this burst, the authors estimate a $\theta_{\mathrm{jet}}$
of 1\arcdeg--2\arcdeg; a similar collimation angle of $\sim1.6\arcdeg$
can be roughly estimated for GRB~130505A from the latest break in its XRT light curve at $\sim$29~ks,
as suggested by the automated analysis.

\section{SUMMARY AND CONCLUSIONS}

We have presented the IPN localization and the multi-wavelength study
of \GRB, the brightest long GRB detected by KW during its almost 19 years of observations.

The \GRB\, prompt emission is characterized by an unprecedented photon flux, whose
fast evolution is accompanied by the equally drastic decay of the peak energy of the spectrum.
At z$\approx$1, the huge energy flux measured by \KW\, implies equally enormous values of
the isotropic-equivalent energy in the source frame, making \GRB\, one of the most energetic
and the most luminous $\gamma$-ray burst observed since the beginning of the cosmological era in 1997.
At the same time we found the key spectral and temporal characteristics of the prompt emission
to be in the range typical of other GRBs, suggesting that the same progenitor model for most
long bursts is also applicable to \GRB.
Having a moderate intrinsic peak energy of the time-integrated spectrum and a typical duration,
the burst is close to being a bright outlier of both the Amati and Yonetoku relations;
this favors a hypothesis that the huge isotropic-equivalent energy
and luminosity result, among other reasons, from a highly collimated emission.

A tail of the soft $\gamma$-ray emission was detected out to $\sim$700~s after the trigger,
with temporal and spectral behavior suggesting an early, bright $\gamma$-ray afterglow;
these observations partially fill a gap between the prompt emission and the \SW\, observations,
which started $\sim1.2$~d later. We found the brightness of the early $\gamma$-ray and the late-time
X-ray afterglows to be in reasonable agreement with reported prompt-afterglow luminosity correlations.
This consistency suggests that the mechanism and efficiency of energy transfer from the \GRB\,
central engine to the blast wave and then to the surrounding medium are fundamentally no different 
from those inherent to most GRBs.

\SW/XRT and \SW/UVOT observed the bright afterglow until 48 days after the burst
and revealed a steady decline of the X-ray and UV/optical flux with no evidence of a jet break.
Our analysis of the multi-wavelength data suggests the post-break scenario for the \SW-observed afterglow,
with a hard underlying electron spectrum and ISM-like circum-burst environment implied.
From the combined light curve, which incorporates the extrapolated early $\gamma$-ray afterglow
and the late-time XRT observations, we estimate the jet break time to be 0.2--1.2~days. 
The small implied collimation angle of 1.7$\arcdeg$-3.4$\arcdeg$ results in reasonable values of
the corrected radiated energy and the peak luminosity.
We note, however, that, among such tightly collimated events, both $E_{\gamma}$ and $L_{\gamma}$
for \GRB\, are on the upper edges of their distributions; this stresses that the burst's intrinsic
brightness is still outstanding. We conclude, therefore, that the enormous luminosity of \GRB\,
is the result of extreme values for several defining characteristics, in which a highly collimated
emission must play a key role.

Finally, we estimate a detection horizon for a similar ultraluminous GRB of z$\sim$7.5 for \KW\,
and z$\sim$12 for \SW/BAT, which emphasizes the importance of $\gamma$-ray bursts as probes
of the early Universe.

.

The \KW\, experiment is supported by a Russian Space Agency contract, RFBR
grants 12-02-00032a and 13-02-12017 ofi-m. KH acknowledges support from
NASA grants NNX07AR71G (MESSENGER Participating Scientist Program)
and NNX12AE41G (Astrophysics Data Analysis Program).
SRO acknowledges support from the UK Space Agency.
VM and BS acknowledge financial contribution from the agreement ASI-INAF I/009/10/0.

This work made use of data supplied by the UK Swift Science Data Centre at the University of Leicester.


%

%

\end{document}